\documentclass[mathpazo,pdftex]{cicp}

\usepackage[utf8]{inputenc}
\usepackage[T1]{fontenc}
\usepackage{amsmath,amssymb,amsfonts,bm,empheq}
\usepackage{etoolbox}
\usepackage[ruled,linesnumbered]{algorithm2e}
\usepackage{hyperref}

\usepackage{changes}
\setdeletedmarkup{{\color{black}\sout{#1}}}

\newcommand{\inner}[2]{\langle#1,#2\rangle}

\renewcommand{\d}{\ensuremath{\mathrm{d}}}
\newcommand{\dt}{ \ensuremath{\mathrm{d} t } }
\newcommand{\dx}{ \ensuremath{\mathrm{d} x} }

\begin{document}
\title{Generating Samples of Stationary Distributions of 
Weakly Interacting  Diffusion Models Without Finite Particle Truncation: A Weak Generative Approach}


\author[Zhiqiang Cai et.~al.]{Zhiqiang Cai\affil{1},
      Chengyu Liu\affil{2}, and Xiang Zhou\affil{3}\corrauth}
\address{\affilnum{1}\ Department of Mathematics, The University of Hong Kong, Pokfulam Road, Hong Kong SAR \\
\affilnum{2}\ Department of Data Science, City University of Hong Kong, Kowloon, Hong Kong SAR \\
\affilnum{3}\ Department of Mathematics, City University of Hong Kong, Kowloon, Hong Kong SAR}
\emails{{\tt zqcai@hku.hk} (Z.~Cai), {\tt cliu687-c@my.cityu.edu.hk} (C.~Liu),
{\tt xiang.zhou@cityu.edu.hk} (X.~Zhou)}


\begin{abstract}
Computing the stationary probability density and generating corresponding samples for the mean-field model of an infinite number of weakly interacting diffusion particles pose significant numerical challenges, particularly in the phase transition regime where the interchangeability of infinite-time and infinite-particle limits breaks down. Traditional approaches, such as direct simulation of finite-particle systems, often fail to accurately pinpoint multiple stationary distributions in the mean-field meta-stable setting. On the other hand, solving the high-dimensional McKean-Vlasov partial differential equation   using neural networks typically yields only the density function, limiting its utility for estimating statistical quantities from  generating samples.
In this work, we propose a novel generative framework based on the weak PDE formulation of the mean-field model to address these challenges. Our approach simultaneously computes the stationary distributions of McKean-Vlasov processes and generates independent and identically distributed samples that satisfy these distributions. This integrated approach not only reveals the true stationary distributions without the random perturbation of finite particle truncation,  but also offers deeper insight into the system’s behavior in the mean‑field limit. 
Extensive numerical experiments demonstrate the effectiveness of the proposed method, showcasing its ability to accurately approximate stationary distributions, capture intricate phase transitions, and handle high-dimensional complex systems. 

\end{abstract}


\maketitle

\section{\label{sec:level1}Introduction}
In $N$-body interacting particle systems, the stochastic evolution of each particle is influenced both by its own state and by the empirical distribution of all particles in the system.
The applications of these interacting stochastic models span a wide range of disciplines, including polymeric fluids \cite{xiang2004,OttingerBook}, granular materials \cite{bolley2013uniform,carrillo2003kinetic}, mathematical biology \cite{keller1971model}, galactic dynamics \cite{binney2011galactic}, synchronization \cite{kuramoto1981rhythms}, and plasma physics \cite{bittencourt2013fundamentals}. In each of these fields, emergent collective behaviors arise at the macroscopic level as a result of mean-field interactions.

The complex collective dynamics of interacting particle systems can be described effectively through the mean‑field limit as $N \to \infty$. Over any fixed time interval $[0,T]$, large‑$N$ systems asymptotically behave as a collection of independent particles—a phenomenon termed ``propagation of chaos” \cite{ChaintronReview2022I,jabin2014review,Sznitman1991}. In this limit, the evolution of the one‑particle (mean‑field) probability density is governed by a nonlinear Fokker–Planck equation \cite{frank2005nonlinear}, which corresponds to a nonlinear stochastic differential equation known as the McKean–Vlasov process \cite{Kac1956-hi,McKean1966,McKean1967}.

The stationary distribution, as the solution of the stationary nonlinear Fokker--Planck equation, represents the equilibrium state of the system as time tends to infinity. Understanding this equilibrium is crucial for analyzing the macroscopic behavior of the mean-field system, as it summarizes the collective dynamics of the interacting particles.
 In a diffusion system described by a free energy—comprising an individual potential, an interaction potential, and an entropy—the stable stationary distributions correspond to local minima of the free energy \cite{carrillo2020long}. Extensive and ongoing research has explored the long‑time behavior of McKean–Vlasov processes, addressing topics such as large deviations, phase transitions, and uniform propagation of chaos; see, for example, \cite{Dawson1983,DW1989,GVALANI2020,jabin2014review} and references therein.

Numerically, two distinctive  approaches exist for computing the stationary distribution of a McKean–Vlasov system. The first is to solve the nonlinear Fokker–Planck equation for the probability density function while incapable of generating iid samples. Classical discretization methods—such as finite‑difference or finite‑element schemes on a mesh—suffer from the curse of dimensionality as the particle dimension grows. 
Additional challenges include ensuring strict positivity and normalization of the numerical density, as well as accurately evaluating the interaction terms.
Using deep neural networks,
such as tensor neural network
\cite{WANG2025107165}
to solve the density function may  overcome the curse of dimensionality, but presents difficulty in 
 efficiently managing the integral for interaction kernel.

A second line of research works  produces samples, but lacks  a feasible method for determining the density function in high dimensions.
This traditional approach  involves the direct 
 Monte Carlo simulation of   $N$ interacting particles over a long time horizon $T$.  This method assumes that the large-$T$ limit of the $N$-particle empirical distribution -- or its one-particle marginal -- converges to the invariant distribution of the corresponding  mean field system. In addition, such simulations are computationally demanding, requiring both a large number of particles ($N \gg 1$) and a long integration time ($T \gg 1$). 
 
A major limitation of particle simulation arises from the subtle issue of interchanging the limits   $N\to\infty$ and $T\to\infty$.
Although the interchange is often justified by time‑uniform propagation of chaos \cite{guillin2024uniform}, it can fail in systems that exhibit phase transitions, where McKean–Vlasov dynamics possess multiple invariant distributions \cite{carrillo2020long,monmarche2025long,PaulEmmanuel2022}.
In such multi‑stable regimes, simulating a finite particle system ( $N$ fixed ) over long times ($T \to \infty$) causes the empirical measure to   randomly transition between meta‑stable states (in the space of probability measures). This occurs because finite‑particle sampling introduces stochastic perturbations on the  order $1/\sqrt{N}$ to the mean‑field dynamics \cite{DW1989},   preventing the  reliable convergence to any local minimum of the free energy\cite{GVALANI2020,tugaut2014phase}.

Therefore, to faithfully compute the density function and generate independent and identically distributed (i.i.d.) samples from the invariant measures of McKean–Vlasov mean‑field systems, there is a clear need for new sampling techniques that are rooted in the mean‑field model -- such as directly addressing the McKean–Vlasov PDE -- rather than relying on finite‑particle simulations. 
Moreover, to support estimation tasks that involve density-based functionals, such as the entropy production $\mathbb{E}_X \log p(X)$ \cite{boffi2024deep, LevyEPR2026PRL}, a sampling approach that inherently yields a density function $p$ is essential to avoid  the need for global density estimation or interpolation in high-dimensional spaces from the  Monte Carlo samples of the stochastic dynamics.

In this work, we  adopt a generative approach  to compute the stationary distributions of McKean-Vlasov mean field model. This generative approach  
facilitates the computation of the  mean-field interaction term   by  using samples from the generative model. This approach employs a neural-network-based transport map to parameterize the target invariant distribution. As in \cite{cai2026weak},  we derive the   weak loss function  for the stationary McKean-Vlasov PDE to  substantially reduce the computational cost  compared to  the  standard least-squares loss.
To overcome the nonlinearity due to the mean field interaction,  we design  two  schemes  which either   directly incorporate  the generative map   into the mean-field interaction or 
employ  a self-contained Picard iteration 
 with a frozen mean-field distribution from the previous iteration.

The main contributions of this work are summarized below:
\begin{enumerate}
    \item We develop a weak generative sampler that constructs a differentiable generative map for  the stationary nonlinear 
     Fokker-Planck equation. This map efficiently generates i.i.d.  samples following the true stationary measure of the McKean-Vlasov system by transforming samples from a simple base distribution.
    
\item   Our methodology operates directly on the mean-field limit and is entirely free from finite-particle approximations. Consequently, the samples generated by our model accurately represent an invariant measure of the underlying McKean-Vlasov process, even when the phase transitions occur.

\item  
The proposed framework is highly general. It accommodates McKean-Vlasov systems with non-gradient drift and interaction terms that possess stationary distributions, without being restricted to gradient systems characterized by a free energy. Furthermore, by employing a parametric normalizing flow, our method is capable of learning invariant measures that assume a parametric form.

\item We provide comprehensive numerical validation across a suite of challenging test cases: meta-stable McKean-Vlasov systems, a mean-field model of active particles with parametric interactions, a non-gradient high-dimensional system with Gauss--Morse interaction and a high-dimensional system with Coulombic interaction. These experiments demonstrate the method's capability to reliably handle complex phenomena   and to accurately learn high-dimensional probability distributions
for the mean-field system.

\end{enumerate}

\par

{\bf Related Works.} To conclude this introduction, we present a selective  review of related works. 
\par

{\it Stationary distribution and phase transition of McKean-Vlasov system:} The mathematical theory concerning the long-time behavior and phase transitions of McKean-Vlasov equations has been  extensively covered in  literature like \cite{carrillo2020long,duerinckx2020mean,GVALANI2020,tugaut2014phase}.  The time uniform propagation of chaos  \cite{guillin2024uniform,monmarche2025long}
 and the  incommutable issue of two limits  of large $T$ and large $N$
are discussed in  \cite{guillin2024uniform,guillin2022uniform}.
For a comprehensive overview, we refer readers to   \cite{ChaintronReview2022I,ChaintronReview2022II,jabin2014review,PaulEmmanuel2022}.

{\it Solving  Fokker-Planck equation  by transport map:}  
 \cite{lu2024score} applies the   score-based transport method   \cite{boffi2023probability,shen2024entropy} to mean-field Fokker-Planck equations with the initial value problems over a finite time horizon $[0,T]$ and does not address the computation of invariant measures.
The generative approach based on the   weak formulation  of Fokker-Planck equations is investigated in  \cite{cai2026weak}. The concept of adaptive training facilitated by a generative map was presented in  \cite{zeng2023adaptive}.

\par 

{\it Approximation by $N$-body interactive particle  system:}
 \cite{li2025solving,li2020random} proposed to sample the $N$-body Gibbs invariant measure using Metropolis-adjusted Langevin dynamics, accelerated by the random batch method  \cite{Jin2020RBM}. The stationary distribution of the McKean-Vlasov equation is then approximated by the one-particle empirical distribution of this finite particle system. A significant limitation of this approach   is its reliance on assumptions like the uniform propagation of chaos to ensure the commutability  of the large $T$ and $N$ limits. Consequently, it is inapplicable to meta-stable systems characterized by multiple mean-field invariant measures. A further practical restriction is its restriction to gradient systems, for which the $N$-body Gibbs measure must be explicitly formulated.

\medskip 
The remainder of this paper is structured as follows.   Section~\ref{sec:2} introduces the foundational background on stochastic interacting particle systems and their mean-field limit, the McKean-Vlasov, and uses the Desai--Zwanzig model to illustrate the limitations of direct simulation under phase transitions..
In Section~\ref{sec:3}, we detail the application of the   Weak Generative Sampler (WGS) framework to 
the McKean--Vlasov setting and present the two proposed numerical schemes for treating the nonlinear interaction term.   Section~\ref{sec:4} is devoted to numerical experiments that validate the efficacy and performance of our methods. Finally, Section~\ref{sec:5} offers concluding remarks. The appendix includes 
some details of a bi-stable McKean-Vlasov model.

\section{Stochastic Interactive Particle Systems and Phase Transitions}\label{sec:2}
\subsection{Stochastic Interactive Particle Systems}

We consider a system of \(N\) interacting particles in \(\mathbb{R}^d\) governed by the coupled stochastic differential equations
\begin{equation}\label{eqn:ia_sde}
    \mathrm{d}X_i
    = f(X_i)\,\mathrm{d}t
    - \frac{1}{N} \sum_{j \neq i} K(X_i, X_j)\,\mathrm{d}t
    + \sqrt{2\epsilon}\,\mathrm{d}\mathbf{B}_t^i,
    \qquad i = 1,\dots,N,
\end{equation}
where \(\{\mathbf{B}_t^i\}_{i=1}^N\) are independent standard \(d\)-dimensional Brownian motions. Here \(X_i \in \mathbb{R}^d\) denotes the position of the \(i\)-th particle, \(f:\mathbb{R}^d \to \mathbb{R}^d\) is a deterministic external force, and \(K:\mathbb{R}^d \times \mathbb{R}^d \to \mathbb{R}^d\) is an interaction kernel.
Define the global drift field \(F_N:(\mathbb{R}^d)^N \to (\mathbb{R}^d)^N\) componentwise by
\[
F_{N,i}(\mathbf{x}_1,\dots,\mathbf{x}_N)
:= f(\mathbf{x}_i) - \frac{1}{N} \sum_{j \neq i} K(\mathbf{x}_i,\mathbf{x}_j),
\qquad i=1,\dots,N.
\]
Let \(p_{N,t}=p_N(t,\mathbf{x}_1,\dots,\mathbf{x}_N)\) denote the joint density of \((X_1,\dots,X_N)\). Then \(p_{N,t}\) satisfies the \(N\)-particle forward Kolmogorov equation
\[
\partial_t p_{N,t}
= \mathcal{L}_N p_{N,t}
:= -\nabla_{\mathbf{x}} \cdot \bigl(F_N p_{N,t}\bigr)
   + \epsilon \Delta_{\mathbf{x}} p_{N,t}.
\]

The scaling \(1/N\) ensures that the interaction experienced by each particle remains of order one as \(N\to\infty\), and thus places the system in the mean-field regime. In the limit \(N\to\infty\), the empirical measure is expected to converge to a deterministic law \(\bar p_t=\mathrm{Law}(\bar X_t)\), where \(\bar X_t\) solves the McKean--Vlasov SDE
\begin{equation}\label{eqn:mvsde}
    \mathrm{d}\bar X_t
    = f(\bar X_t)\,\mathrm{d}t
    - (K * \bar p_t)(\bar X_t)\,\mathrm{d}t
    + \sqrt{2\epsilon}\,\mathrm{d} \mathbf{B}_t,
\end{equation}
where $\{\mathbf{B}_t\}_{t \geq 0}$ is a standard $d$-dimensional Brownian motion. The convolution operator $*$ is defined by $
(h * \phi)(\mathbf{x}) := \int_{\mathbb{R}^d} h(\mathbf{x}, \mathbf{y})\phi(\mathbf{y})\,\mathrm{d}\mathbf{y},$ 
and $\bar{p}_t: \mathbb{R}^d \rightarrow \mathbb{R}_{\geq 0}$ denotes the probability density function of $\bar{X}_t$. $\bar{p}_t$ satisfies 
the following McKean-Vlasov equation -- a {\it nonlinear} Fokker--Planck equation:
\begin{equation}\label{eqn:nfpe}
    \partial_t \bar p_t = \mathcal{L}_{\bar p_t} \bar p_t,
\end{equation}
where
\begin{equation}\label{eqn:ndo}
    \mathcal{L}_q p
    := -\nabla \cdot \bigl(p(f-K*q)\bigr) + \epsilon \Delta p.
\end{equation}

Under suitable assumptions, one has propagation of chaos: for each fixed \(k \geq 1\) and \(T>0\),
\[
\lim_{N\to\infty}
\sup_{0\le t\le T}
\bigl\|p_{N,t}^{(k)}-\bar p_t^{\otimes k}\bigr\| = 0\]
in an appropriate topology, for instance weak convergence or a Wasserstein distance. If the convergence holds uniformly for all \(t\ge 0\), that is, with \(\sup_{t\ge 0}\) in place of \(\sup_{0\le t\le T}\), then one speaks of time-uniform propagation of chaos.

An important subclass is given by gradient systems, for which
\[
f(\mathbf{x})=-\nabla V(\mathbf{x}),
\qquad
K(\mathbf{x},\mathbf{y})=\nabla_{\mathbf{x}}W(\mathbf{x}-\mathbf{y}).
\]
Assume that the interaction potential \(W\) is even. Then the \(N\)-particle energy is
\begin{equation}
\mathcal{E}_N(\mathbf{x}_1,\ldots,\mathbf{x}_N)
= \sum_{n=1}^N V(\mathbf{x}_n)
+ \frac{1}{2N}\sum_{i\ne j} W(\mathbf{x}_i-\mathbf{x}_j),
\end{equation}
and the corresponding Gibbs measure
\begin{equation}
    \pi_N(\mathbf{x}_1,\ldots,\mathbf{x}_N)
    \propto
    \exp\left(
    -\frac{1}{\epsilon}\mathcal{E}_N(\mathbf{x}_1,\ldots,\mathbf{x}_N)
    \right)
\end{equation}
is invariant for the \(N\)-particle dynamics. In the mean-field limit, the McKean--Vlasov equation \eqref{eqn:nfpe} is the Wasserstein gradient flow of the free-energy functional \(\mathcal{F}\) \cite{Sznitman1991,JKO1998},
\begin{equation}
\begin{split}
\mathcal{F}[p]
&=
\int_{\mathbb{R}^d} V(\mathbf{x})\,p(\mathbf{x})\,\mathrm{d}\mathbf{x}
+\frac{1}{2}
\iint_{\mathbb{R}^d\times\mathbb{R}^d}
W(\mathbf{x}-\mathbf{y})\,p(\mathbf{x})\,p(\mathbf{y})\,
\mathrm{d}\mathbf{x}\,\mathrm{d}\mathbf{y} 
+\epsilon\int_{\mathbb{R}^d} p(\mathbf{x})\log p(\mathbf{x})\,\mathrm{d}\mathbf{x}.
\end{split}
\end{equation}
Accordingly, under suitable regularity and integrability assumptions, stationary solutions of \eqref{eqn:nfpe} coincide with critical points of \(\mathcal{F}\) on \(\mathcal{P}(\mathbb{R}^d)\).

For non-convex confining and/or interaction potentials and sufficiently small noise level \(\epsilon\), the free energy \(\mathcal{F}\) may admit multiple local minimizers. In that regime, the solution \(\bar p_t\) of the McKean--Vlasov equation may converge, depending on the initial distribution \(\bar p_0\), to different stationary states corresponding to different local minimizers of \(\mathcal{F}\).

A basic consequence is that accumulation points of the one-particle marginals \(p_N^{(1)}\) of the \(N\)-particle Gibbs measures need not coincide with the stationary states selected by the McKean--Vlasov dynamics. Thus the mean-field limit \(N\to\infty\) and the long-time limit \(t\to\infty\) need not commute. If time-uniform propagation of chaos is available, then the large-\(N\) limit of the invariant one-particle marginals agrees with the stationary state selected by the McKean--Vlasov dynamics, thereby justifying the interchange of the limits \(N\to\infty\) and \(t\to\infty\). For non-convex potentials, however, establishing this equivalence a priori is in general highly nontrivial. This is one reason to work directly with the McKean--Vlasov equation rather than with the \(N\)-particle approximation.

\subsection{Phase transition in the Desai--Zwanzig model}

To illustrate the emergence of phase transitions in McKean--Vlasov dynamics, we consider the classical Desai--Zwanzig model \cite{DesaiZwanzig1978}. In one dimension, the confining potential is
$
V(x)=(x^2-1)^2,
$
and the interaction kernel is
$
W(x-y)=\vartheta (x-y)^2/2,
$
where $\vartheta>0$ denotes the interaction strength. The corresponding nonlinear mean-field dynamics are
\[
\mathrm{d}X_t
=
-\bigl(V'(X_t)+\vartheta(X_t-m_t)\bigr)\,\mathrm{d}t
+
\sqrt{2\epsilon}\,\mathrm{d}B_t,
\qquad
m_t=\mathbb{E}[X_t].
\]

For each $m\in\mathbb{R}$, define
\[
\rho_m(x)
=
\frac{1}{Z_m}\exp\!\left(-\frac{1}{\epsilon}\Bigl[(x^2-1)^2+\frac{\vartheta}{2}(x-m)^2\Bigr]\right),
\]
where $Z_m$ is the normalization constant. The stationary solutions are characterized by the fixed-point relation
\[
m=\int_{\mathbb{R}} x\,\rho_m(x)\,\mathrm{d}x.
\]
Since $V$ is even, $m=0$ is always a solution. Moreover, it is well known that this model undergoes a pitchfork bifurcation: for $\epsilon=1$, there exists a critical value $\vartheta_c\approx 1.858$ such that for $\vartheta<\vartheta_c$ the stationary solution is unique, while for $\vartheta>\vartheta_c$ two additional stable asymmetric stationary solutions appear; see \cite{DesaiZwanzig1978}. Thus, in the supercritical regime, the stationary McKean--Vlasov equation admits multiple solutions, including an unstable symmetric branch.

\begin{figure}[ht]
    \centering
    \includegraphics[scale=0.19]{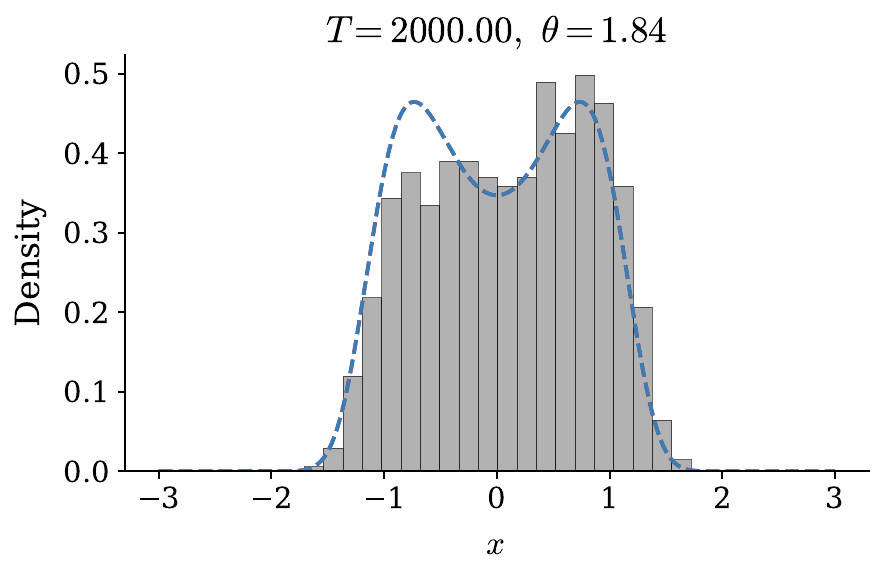}
    \includegraphics[scale=0.19]{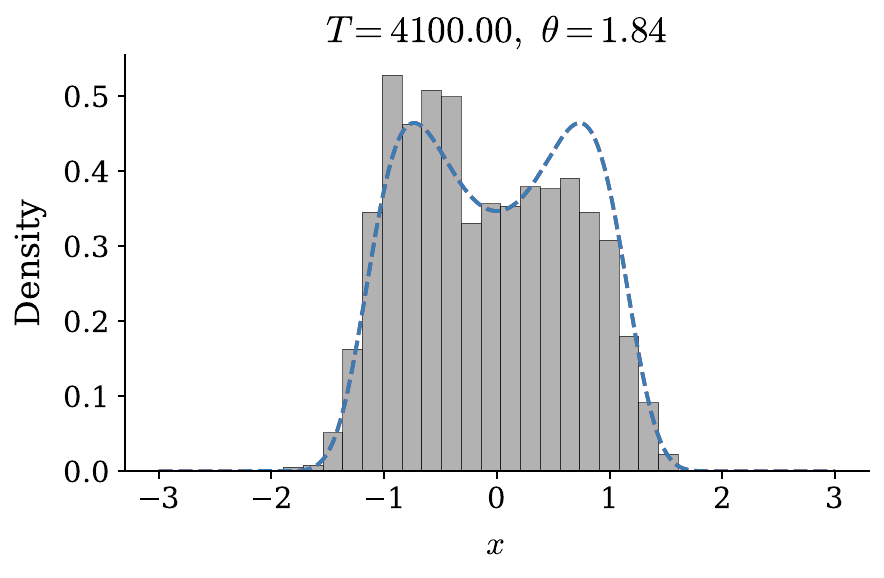}
    \includegraphics[scale=0.19]{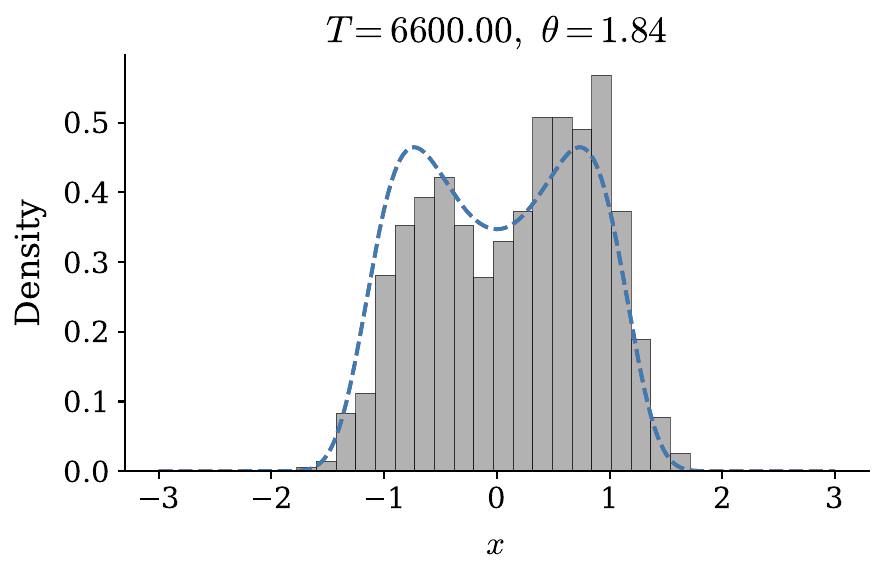}
    \includegraphics[scale=0.19]{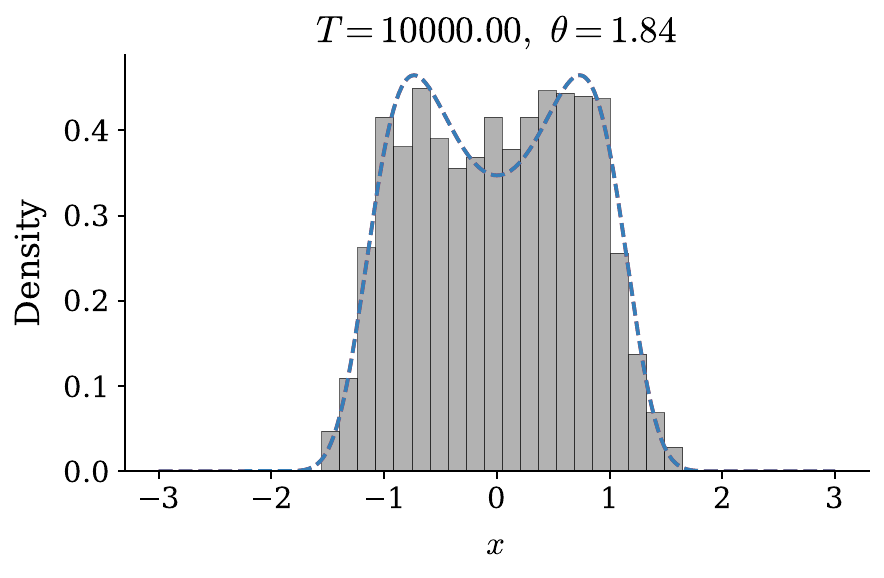}
    \includegraphics[scale=0.19]{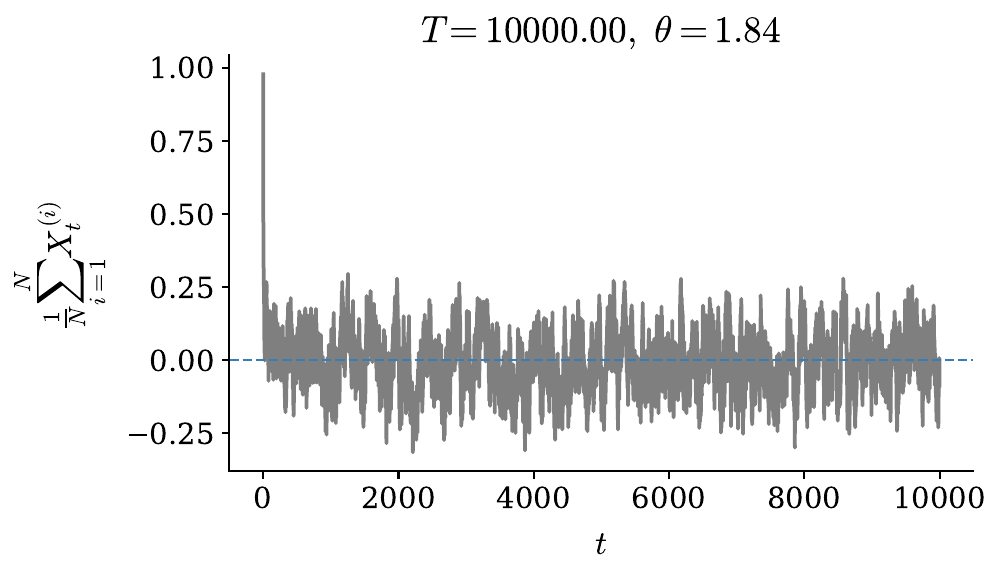}
    
    \includegraphics[scale=0.19]{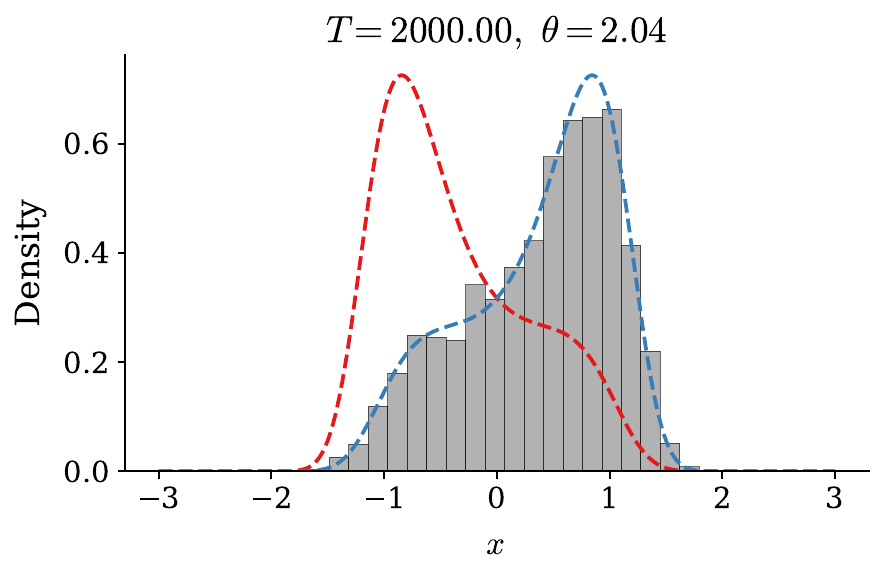}
    \includegraphics[scale=0.19]{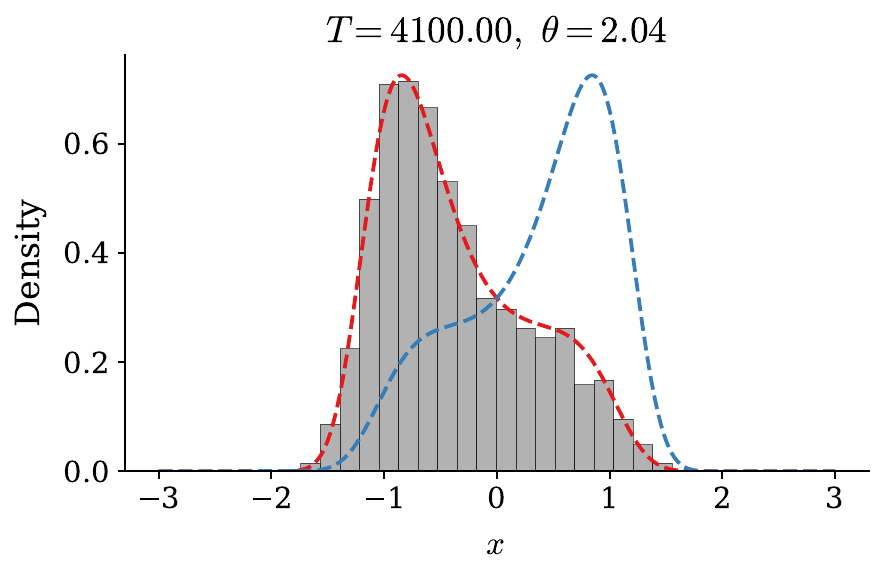}
    \includegraphics[scale=0.19]{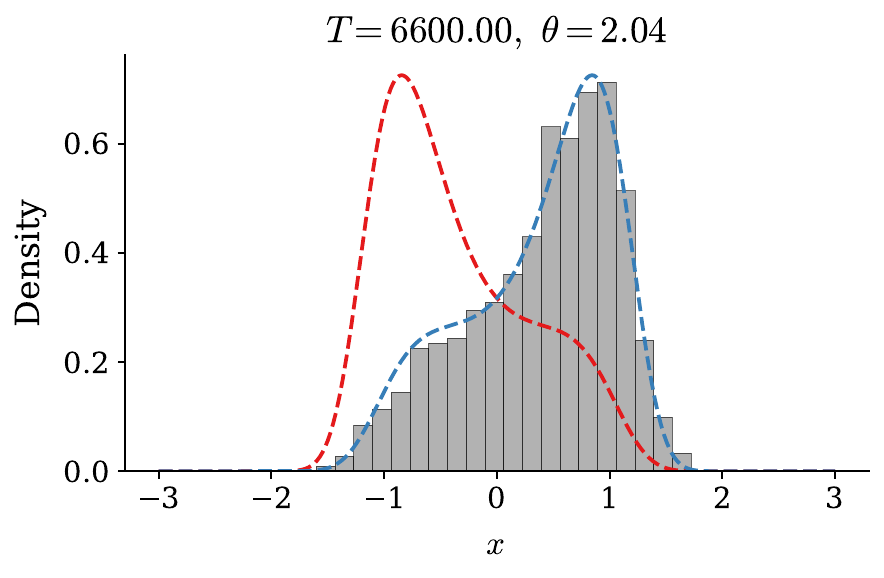}
    \includegraphics[scale=0.19]{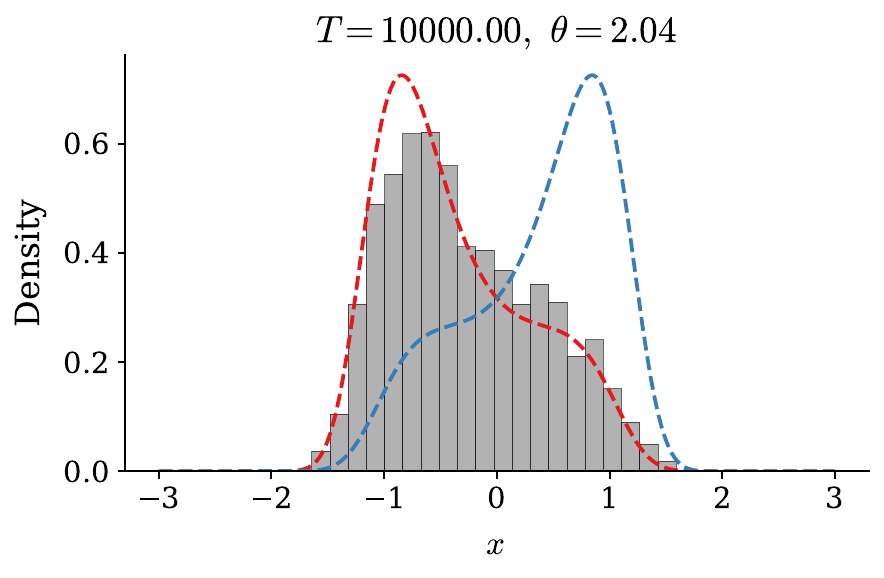}
    \includegraphics[scale=0.19]{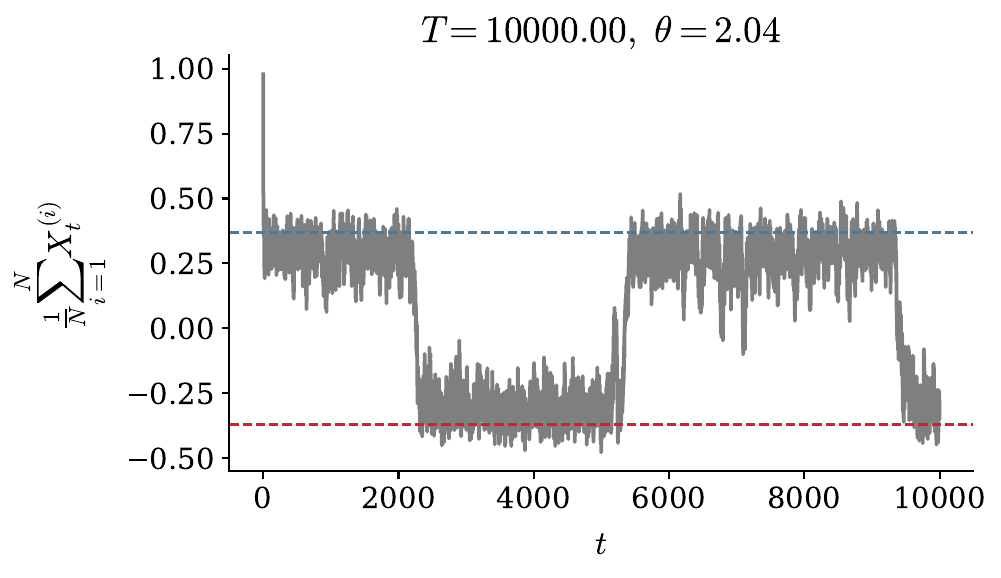}
    
    \caption{Empirical distributions of the particle positions at \(t=2000, 4100, 6600,\) and \(10000\), together with the corresponding empirical mean trajectories, for the Desai--Zwanzig model with \(N=2000\). The particle system is simulated by the Euler--Maruyama scheme with time step \(\Delta t=0.01\) up to final time \(T=10^4\). The top row corresponds to \(\vartheta=1.84<\vartheta_c\), where the empirical distribution remains close to a symmetric profile and relaxes toward the unique symmetric stationary state; the rightmost panel shows that the empirical mean fluctuates around zero. The bottom row corresponds to \(\vartheta=2.04>\vartheta_c\), where the system is in the supercritical regime and the empirical distribution becomes asymmetric, reflecting the presence of two stable symmetry-broken stationary branches; the rightmost panel shows that the empirical mean spends long time intervals near two nonzero values and occasionally switches between them due to finite-particle effects. The dashed horizontal lines indicate the theoretical stationary mean values obtained from the fixed-point equation.}
    \label{fig:desai_zwanzig_phase_transition}
\end{figure}

This behavior is illustrated in Figure~\ref{fig:desai_zwanzig_phase_transition}. The first two rows show the empirical distributions of the particle positions at several fixed times for $\vartheta=1.84$ and $\vartheta=2.04$, respectively, with $N=2000$, starting from the same initial distribution $\mathcal{N}(1,0.2)$. Since $\vartheta=1.84<\vartheta_c$, the empirical distribution remains close to a symmetric profile, and the empirical mean fluctuates around zero, consistent with convergence toward the unique symmetric stationary state. By contrast, for $\vartheta=2.04>\vartheta_c$, the system is in the supercritical regime and the empirical distribution becomes asymmetric, reflecting the influence of the two stable symmetry-broken stationary branches. The corresponding empirical mean no longer stays near zero, but spends long time intervals near two nonzero values and occasionally switches between them due to finite-particle fluctuations. 
This metastable switching behavior shows that direct particle simulation does not by itself provide a complete numerical characterization of the stationary solutions of the mean-field equation. In particular, in the supercritical regime, a single trajectory does not converge to the unstable symmetric branch and only samples the stable symmetry-broken states selected by the finite-particle dynamics.

\section{Main method}\label{sec:3}
In this section, we shall develop the mean field version of the weak generative approach to find the stationary measures for the mean field McKean-Vlasov system. 
We first introduce  the \emph{weak generative sampler} (WGS) 
for the classic Fokker-Planck equation \cite{cai2026weak} in Section \ref{sec:3.1}.  The new schemes for the McKean-Vlasov equation are proposed to probe the stationary  measures to resolve the nonlinearity issue due to  interaction effect in Section \ref{sec:3.2}. Section~\ref{sec:3.3}  includes  some details on  the
network architecture for the generative map,
and presents  how to extend our method  to the  system  in a parametric form to demonstrate the generalization capacity .

\subsection{Weak Generative Approach  for Invariant Measure of SDE}\label{sec:3.1}
The generative approach is to  use a transport map $G_{\bm{\theta}}$, parameterized by ${\bm{\theta}}$,  to map a base distribution $\rho_{\text{base}}$ (e.g., Gaussian or uniform) to a target distribution $p$, meaning that  
if a random variable $\mathbf{z} \sim \rho_{\text{base}}$, then the transported random variable  $\mathbf{x} = G_{\bm{\theta}}(\mathbf{z})$ has the distribution $p$. 
In general, $G_{\bm{\theta}}$ is a diffeomorphism so that the density $p(\mathbf{x})$ can be theoretically computed via the formula of change-of-variable: \begin{equation} \label{283}    p(\mathbf{x})=\rho_{\text{base}}\left(\mathbf{z}\right) \cdot\left|\operatorname{det}\left(\nabla G_{\bm{\theta}}(\mathbf{z})\right)\right|^{-1},
    \quad  \mathbf{z}=G_{\bm{\theta}}^{-1}(\mathbf{x}).
\end{equation}

For the linear stationary Fokker--Planck equation $\mathcal{L}p=0$, defined by $\mathcal{L}p:=-\nabla\cdot (pf) +\epsilon\Delta p$, the conventional mean-square loss employed in   physics-informed neural networks is
$$\int (\mathcal{L} p(\mathbf{x}))^2 p(\mathbf{x})\dx =\mathbb{E}_{X\sim p}(\mathcal{L} p(\mathbf{x}))^2  $$ 
where the adaptive sampling distribution for   training is set as $p$ itself \cite{tang2022adaptive} and the optimization is usually conducted sequentially by freezing the sampling density in each iteration. However, the differential operator involving  the unknown density function, as expressed in \eqref{283}, imposes a substantial computational cost.

The weak loss first proposed in \cite{cai2026weak} is  to   leverage the \emph{weak formulation} of the stationary Fokker--Planck equation   by applying a set of test functions $\varphi$ so that  
\begin{equation}\label{eqn:sfpe}
\inner{\varphi}{\mathcal{L} p}=0, \quad \forall \varphi \in C_c^\infty(\mathbb{R}^d),
\end{equation}
which is further  rewritten as 
\begin{equation}\label{eqn:wfpe}
\inner{\mathcal{L}^* \varphi}{p} = 0, \quad \forall \varphi \in C_c^\infty(\mathbb{R}^d),
\end{equation}
where $\mathcal{L}^*$ is the adjoint operator  of $\mathcal{L}$.
This method addresses this system of equations \eqref{eqn:wfpe} through a probabilistic approach by assigning a non-degenerate probability distribution $\mathbb{P}$ for the test functions and  cast the problem as the following 
\begin{equation}\label{eqn:wsfpe_e}
\min_{p} \mathbb{E}_{\varphi \sim \mathbb{P}} \Big[\mathbb{E}_{\mathbf{x} \sim p}\big[\mathcal{L}^* \varphi(\mathbf{x})\big]\Big]^2.
\end{equation}
With the aid of the generative model, the sample   $x = G_{\bm{\theta}}(z)$, then the optimization problem becomes
\begin{equation}\label{eqn:eloss}
\min_{G_{\bm{\theta}}} \mathbb{E}_{\varphi \sim \mathbb{P}} \bigg[\mathbb{E}_{\mathbf{z} \sim \rho_{\text{base}}} \Big[\mathcal{L}^* \varphi\big(G_{\bm{\theta}}(\mathbf{z})\big)\Big]\bigg]^2.
\end{equation}

The set of test functions, $\varphi_\zeta$, are  constructed using Gaussian kernels centered at data points $\zeta\in\mathbb{R}^d$:
\begin{equation}\label{eqn:test_function}
\varphi_\zeta(\mathbf{x}) = \exp\left(-\frac{\|\mathbf{x} - \zeta\|_2^2}{2\kappa^2}\right),
\end{equation}
where $\kappa$ is a hyperparameter determining the kernel width.  $\zeta$ are adaptively chosen from the samples in the previous iteration. More specific details can be found in \cite{cai2026weak}.

\subsection{Weak Generative Sampler for  McKean-Vlasov Processes}\label{sec:3.2}
  
In the McKean-Vlasov model, the stationary distribution $p$ is governed by the following nonlinear PDE involving the operator $\mathcal{L}_p$ defined in~\eqref{eqn:ndo}:
$$
\mathcal{L}_p p = 0.
$$

To apply the weak generative framework, we consider the weak form of \eqref{eqn:nfpe}. Let $\varphi \in C_c^\infty(\mathbb{R}^d)$ be a test function, the corresponding adjoint operator of $\mathcal{L}_p$ in $L^2(\mathbb{R}^d)$ is given by:
\begin{equation}\label{eqn:adjoint_dno}
    \mathcal{L}_p^* \varphi := (f - K * p) \cdot \nabla \varphi + \epsilon \Delta \varphi.
\end{equation}
The presence of the convolution term $K * p$ introduces nonlinearity into the operator. Consequently, the core challenge  in this context is the efficient and stable evaluation of the interaction term $K * p$ during optimization.
In the following, we introduce two strategies for implementing WGS in the McKean--Vlasov setting:
\begin{itemize}
\item  The implicit iteration that plugs the generative map $G_{\bm{\theta}}$ directly into the interaction term;
\item The Picard iteration that decouples the generative map $G_{\bm{\theta}}$ from the convolution term by using a frozen distribution   from the previous step for the interaction.   
\end{itemize}
These two approaches offer different trade-offs in terms of stability, convergence, and implementation complexity.

\subsubsection{Implicit iteration}\label{sec:3.2.1}
The straightforward method to handle the interaction term in $\mathcal{L}_p^*$ is to fully use the power of 
deep learning by directly putting the training generative map $G_{\bm{\theta}}$ into the loss function. Formally, we can 
obtain the optimal generative map $G_{{\bm{\theta}}^*}$ by
\begin{equation}\label{eqn:loss_pi}
G_{{\bm{\theta}}^*}=\mathop{\arg\min}_{\bm{\theta}} ~\mathbb{E}_{\varphi\sim\mathbb{P}}\bigg[\mathbb{E}_{\mathbf{z}\sim \rho_{\text{base}}}\Big[\mathcal{L}_{p_{\bm{\theta}}}^*\varphi\big(G_{\bm{\theta}}(\mathbf{z})\big)\Big]\bigg]^2.
\end{equation}

In Implicit iteration, the generative map $G_{\bm{\theta}}$ is implicit in the loss function, and then the loss function is nonlinear with respect to $p_{\bm{\theta}}$. Empirically, the first expectation w.r.t $\mathbf{z}\sim\rho_{\text{base}}$ can be approximated by
$$
\frac{1}{N}\sum_{i=1}^N \hat{\mathcal{L}}_{p_{\bm{\theta}}}^*\varphi_j\big(G_{\bm{\theta}}(\mathbf{z}_i)\big),
$$
where
$$
\begin{aligned}
&\hat{\mathcal{L}}^*_{p_{\bm{\theta}}}\varphi_j\big(G_{\bm{\theta}}(\mathbf{z}_i)\big)=\epsilon\Delta \varphi_j\left(G_{\bm{\theta}}(\mathbf{z}_i)\right)+
\left(f\big(G_{\bm{\theta}}(\mathbf{z}_i)\big)-
\frac{1}{N}\sum_{k=1,k \neq i}^N K\left(G_{\bm{\theta}}(\mathbf{z}_i),G_{\bm{\theta}}(\mathbf{z}_k)\right)\right)\cdot\nabla \varphi_j\left(G_{\bm{\theta}}(\mathbf{z}_i)\right).\\
\end{aligned}
$$

We follow the same choice of the test functions in  \cite{cai2026weak} as 
the family of the Gaussian kernel functions. Therefore, the empirical loss function of the Implicit iteration  is
\begin{equation}\label{eqn:iiloss}
    L_I(G_{\bm{\theta}})=\frac{1}{N_\varphi}\sum_{j=1}^{N_\varphi}\left[\frac{1}{N}\sum_{i=1}^N \hat{\mathcal{L}}_{p_{\bm{\theta}}}^*\varphi_j\big(G_{\bm{\theta}}(\mathbf{z}_i)\big)\right]^2+ L_{b}.
\end{equation}
where $L_b$ is the penalty preventing the map from going  to infinity, defined as 
$$
L_b =  \frac{\lambda}{N} \sum_{i=1}^{N} \text{Sigmoid}\left(c\left(\|G_\theta(\mathbf{z}_i) - \mathbf{x}_0\|_2^2 - r^2\right)\right).
$$
Here \(\lambda\), \(r\), and \(c\) are positive hyperparameters, and \(\text{Sigmoid}(x) = 1 / (1 + e^{-x})\). This term imposes a large penalty for samples outside \(B_r(\mathbf{x}_0)\), with \(c\) and \(\lambda\) controlling the strength of the confinement. See Algorithm \ref{alg:wgs_ii}.

\begin{algorithm}[htbp]
\footnotesize
\caption{Implicit iteration for WGS (II-WGS)}\label{alg:wgs_ii}
\SetKwInOut{Input}{Input}\SetKwInOut{Output}{Output}
\Input{Initial generative map $G_{\bm{\theta}}$, the base distribution $\rho_{\text{base}}$; the  hyper-parameters $\gamma>0$, $\kappa>0$, $\lambda>0$, $r>0$, $c>0$. }

\For{$n=1:N_I$}{Sample $\{\mathbf{z}_i\}_{i=1}^N$ from  $\rho_{\text{base}}$\;
Obtain $\{\mathbf{x}_i\}_{i=1}^N$ by $\mathbf{x}_i=G_{\bm{\theta}}(\mathbf{z}_i)$\;
Randomly choose $N_\varphi$ numbers from $1:N$ as index $ind$\;
Split $ind$ into mini-batches of size $N_\varphi^b$\;
\For{$m=1: \lceil N_\varphi/N_\varphi^b \rceil $}{
Obtain  $\{\mathbf{x}_{(j)}\}_{j=1}^{N_\varphi^b}$ by $\mathbf{x}_{(j)}=\mathbf{x}_{ind(m,j)}+\gamma \mathcal{N}(\mathbf{0},\boldsymbol{I}_d)$\;
Construct the test function $\varphi_j$ by Gaussian kernel as
$$
\varphi_j(\mathbf{x})=\exp{\left(-\frac{1}{2\kappa^2}(\mathbf{x}-\mathbf{x}_{(j)})^\top(\mathbf{x}-\mathbf{x}_{(j)})\right)},
$$

Compute the Loss function \eqref{eqn:iiloss}\;
Update the parameters ${\bm{\theta}}$ using the Adam optimizer with a learning rate $\eta$\;
}
}
\Output{The trained generative map $G_{\bm{\theta}}$}
\end{algorithm}

\subsubsection{Picard iteration}\label{sec:3.2.2}
Unlike the Implicit iteration, Picard iteration proposes to avoid the nonlinearity related to $p_{\bm{\theta}}$ that appears in the loss function. The Picard iteration involves utilizing an approximation $\hat{p}$ of $p$ to linearize the nonlinear differential operator in \eqref{eqn:adjoint_dno} for each iteration.
 Formally, let $n$ be the $n$-th iteration following the gradient descent of the loss function. Then, for $(n+1)$-th iteration, the empirical loss function is computed via
\begin{equation}\label{eqn:piloss}
    L_P(G_{\bm{\theta}})=\frac{1}{N_\varphi}\sum_{j=1}^{N_\varphi}\left[\frac{1}{N}\sum_{i=1}^N \hat{\mathcal{L}}_{p^n_{\tilde{{\bm{\theta}}}}}^*\varphi_j\big(G_{\bm{\theta}}(\mathbf{z}_i)\big)\right]^2+ L_{b},
\end{equation}
where
$$
\begin{aligned}
\hat{\mathcal{L}}^*_{p^n_{\tilde{{\bm{\theta}}}}}\varphi_j\big(G_{\bm{\theta}}(\mathbf{z}_i)\big)=\epsilon\Delta \varphi_j\left(G_\theta(\mathbf{z}_i)\right)+
\left(f\big(G_{\bm{\theta}}(\mathbf{z}_i)\big)-\frac{1}{N}\sum_{k=1, k \neq i}^N K\left(G_{\bm{\theta}}(\mathbf{z}_i),G_{\tilde{{\bm{\theta}}}}^n(\mathbf{z}_k)\right)\right)\cdot&\nabla \varphi_j\left(G_{\bm{\theta}}(\mathbf{z}_i)\right).
\end{aligned}
$$
and $p^n_{\tilde{{\bm{\theta}}}}$ denotes the output from the $n$-th gradient descent (For $n=0$, we set the $G_{\tilde{{\bm{\theta}}}}^0$ as same as the initial generative map $G_{\bm{\theta}}$). The detail is in Algorithm \ref{alg:wgs_pi}.

\begin{algorithm}[htbp]
\footnotesize
\caption{Picard iteration for WGS (PI-WGS)}\label{alg:wgs_pi}
\SetKwInOut{Input}{Input}\SetKwInOut{Output}{Output}
\Input{Initial generative map $G_{\bm{\theta}}$, the base distribution $\rho_{\text{base}}$; the  hyper-parameters $\gamma>0$, $\kappa>0$, $\lambda>0$, $r>0$, $c>0$. }

Initialize: $G_{\tilde{{\bm{\theta}}}}^0 = G_{\bm{\theta}}$\;
\For{$n=1:N_I$}{Sample $\{\mathbf{z}_i\}_{i=1}^N$ from  $\rho_{\text{base}}$\;
Obtain $\{\mathbf{x}_i\}_{i=1}^N$ by $\mathbf{x}_i=G_{\bm{\theta}}(\mathbf{z}_i)$\;
Randomly choose $N_\varphi$ numbers from $1:N$ as index $ind$\;
Split $ind$ into mini-batches of size $N_\varphi^b$\;
\For{$m=1: \lceil N_\varphi/N_\varphi^b \rceil $}{
Obtain  $\{\mathbf{x}_{(j)}\}_{j=1}^{N_\varphi^b}$ by $\mathbf{x}_{(j)}=\mathbf{x}_{ind(m,j)}+\gamma \mathcal{N}(\mathbf{0},\boldsymbol{I}_d)$\;

Construct the test function $\varphi_j$ by Gaussian kernel as
$$
\varphi_j(\mathbf{x})=\exp{\left(-\frac{1}{2\kappa^2}(\mathbf{x}-\mathbf{x}_{(j)})^\top(\mathbf{x}-\mathbf{x}_{(j)})\right)},
$$
Compute the loss \eqref{eqn:piloss} using the frozen map $G_{\tilde{\bm{\theta}}}^{n-1}$\;
Update the parameters ${\bm{\theta}}$ using the Adam optimizer with a learning rate $\eta$\;
}
Update: $G_{\tilde{\bm{\theta}}}^{n} = G_{\bm{\theta}}$\;
}
\Output{The trained generative map $G_{\bm{\theta}}$}
\end{algorithm}

\subsection{Generative Maps via Normalizing Flows and its Parametric Representation}\label{sec:3.3}
\subsubsection{Normalizing Flows}
In this work, we employ normalizing flows~ \cite{rezende2015variational} as the parameterization framework for constructing the generative map $G_\theta$ used in our proposed methods. Normalizing flows are invertible neural network models that transform a simple base distribution $\rho_{\text{base}}$ (e.g., Gaussian or uniform) into a complex target distribution $p$ through a series of structured mappings. Specifically, the generative map $G_{\bm{\theta}}$ is expressed as a composition of $L$ bijective transformations $\{g_{{\bm{\theta}}_i}\}_{i=1}^L$, such that  
\[
G_{\bm{\theta}} = g_{{\bm{\theta}}_L} \circ g_{{\bm{\theta}}_{L-1}} \circ \cdots \circ g_{{\bm{\theta}}_1},
\]
where ${\bm{\theta}} = ({\bm{\theta}}_1, {\bm{\theta}}_2, \dots, {\bm{\theta}}_L)$ represents the learnable parameters of the model. For a given input $\mathbf{z} \sim \rho_{\text{base}}$, the output $\mathbf{x} = G_{\bm{\theta}}(\mathbf{z})$ is obtained by sequentially applying these transformations:
\[
\mathbf{x}^{(i)} = g_{{\bm{\theta}}_i}(\mathbf{x}^{(i-1)}), \quad i = 1, 2, \dots, L,
\]
with $\mathbf{x}^{(0)} = \mathbf{z}$ and $\mathbf{x}^{(L)} = \mathbf{x}$. These transformations progressively warp the samples from the base distribution $\rho_{\text{base}}$ into those matching the desired target distribution $p$.

To implement $g_{{\bm{\theta}}_i}$, one popular strategy is to adopt the affine coupling layer design from Real NVP  \cite{dinh2017density,papamakarios2021normalizing}. The affine coupling layer splits the input $\mathbf{x}^{(i-1)} \in \mathbb{R}^d$ into two partitions $(\mathbf{x}_1^{(i-1)}, \mathbf{x}_2^{(i-1)}) \in \mathbb{R}^a \times \mathbb{R}^{d-a}$, where $a$ is a hyperparameter that determines the split dimension. The layer then updates only one part of the input (e.g., $\mathbf{x}_1^{(i-1)}$), while leaving the other part (e.g., $\mathbf{x}_2^{(i-1)}$) unchanged:
\[
\mathbf{x}^{(i)} = g_{{\bm{\theta}}_i}(\mathbf{x}_1^{(i-1)}, \mathbf{x}_2^{(i-1)}) = \big(h_i(\mathbf{x}_1^{(i-1)}; \Theta_i(\mathbf{x}_2^{(i-1)})), \mathbf{x}_2^{(i-1)}\big),
\]
where $\Theta_i: \mathbb{R}^{d-a}\mapsto\mathbb{R}^a$ is parameterized by ${\bm{\theta}}_i=({\bm{\theta}}_i^1,{\bm{\theta}}_i^2)$ and the coupling function $h_i:\mathbb{R}^d \to \mathbb{R}^a$ is defined as:
\[
\begin{aligned}
&h_i\big(\mathbf{x}_1^{(i-1)}; \Theta_i(\mathbf{x}_2^{(i-1)})\big)\\
=& \big(\mathbf{x}_1^{(i-1)} - t_{{\bm{\theta}}^1_i}(\mathbf{x}_2^{(i-1)})\big) \odot \exp\big(-s_{{\bm{\theta}}^2_i}(\mathbf{x}_2^{(i-1)})\big).
\end{aligned}
\]
In the above, $t_{{\bm{\theta}}^1_i}:\mathbb{R}^{d-a} \mapsto \mathbb{R}^a$ and $s_{{\bm{\theta}}^2_i}:\mathbb{R}^{d-a} \mapsto \mathbb{R}^a$ are the translation and scaling functions, respectively, parameterized by neural networks with learnable parameters ${\bm{\theta}}^1_i$ and ${\bm{\theta}}^2_i$. The operator $\odot$ denotes element-wise multiplication.

To ensure all components of the input $\mathbf{z}$ are updated during the transformation, the partition of $\mathbf{x}$ into $(\mathbf{x}_1, \mathbf{x}_2)$ is alternated across successive affine coupling layers. For instance, $\mathbf{x}_2^{(i-1)}$ remains unchanged in one coupling layer but is updated in the subsequent layer. This alternating update scheme increases the flexibility and expressiveness of the model while maintaining computational efficiency.

The overall flexibility of normalizing flows is enhanced by the ability to shuffle the input dimensions or randomize the partitioning strategy. These design choices, combined with the affine coupling layers, allow the generative map $G_{\bm{\theta}}$ to capture a wide range of complex transformations, making it particularly suitable for tasks requiring accurate density estimation and sampling. It is worth noting that continuous normalizing flows (CNFs) could also be considered for constructing the generative map, and their theoretical reliability in sample generation has been recently supported by rigorous convergence analyses~\cite{jing2025convergence}. However, the details about the improvement of network architecture and expressive power are beyond the scope of our work here.

\subsubsection{Parametric Representation for Normalizing Flows}
\label{ssec332}
In practice, our proposed methods can be extended to a family of McKean–Vlasov processes parameterized by certain physical properties, such as temperature or other domain-specific characteristics. Let $\bm{\beta} \in \mathbb{R}^{d^\prime}$ represent the parameter of the McKean–Vlasov process. The training objective for this family of processes can be formulated as the following loss function:
\[
\mathbb{E}_{\bm{\beta} \sim q}\left[L\left(G_{\bm{\theta}}(\bm{\beta})\right)\right],
\]
where $L(G_{\bm{\theta}}(\bm{\beta}))$ is instantiated as either $L_I(G_{\bm{\theta}}(\bm{\beta}))$ for II-WGS or $L_P(G_{\bm{\theta}}(\bm{\beta}))$ for PI-WGS, and $q$ is a uniform distribution over the domain of $\bm{\beta}$. 

To incorporate the parameter $\bm{\beta}$ into the generative map $G_{\bm{\theta}}$ via normalizing flows, the functions $t_{{\bm{\theta}}_i^1}$ and $s_{{\bm{\theta}}_i^2}$ in the affine coupling layer are modified as conditioning networks. Specifically, the transformations $t_{{\bm{\theta}}_i^1}:\mathbb{R}^{d-a+d^\prime} \to \mathbb{R}^a$ and $s_{{\bm{\theta}}_i^2}:\mathbb{R}^{d-a+d^\prime} \to \mathbb{R}^a$ depend explicitly on both $\mathbf{x}_2^{(i-1)}$ and $\bm{\beta}$. Consequently, the coupling function $h_i$ is defined as:
\[
\begin{aligned}
&h_i\big(\mathbf{x}_1^{(i-1)}; \Theta_i(\mathbf{x}_2^{(i-1)}, \bm{\beta})\big) \\
= &\big(\mathbf{x}_1^{(i-1)} - t_{{\bm{\theta}}^1_i}(\mathbf{x}_2^{(i-1)}, \bm{\beta})\big)  \odot \exp\big(-s_{{\bm{\theta}}^2_i}(\mathbf{x}_2^{(i-1)}, \bm{\beta})\big).
\end{aligned}
\]
This formulation allows the parameter $\bm{\beta}$ to be seamlessly coupled into the generative map, enabling the model to adapt to different configurations of the McKean–Vlasov process.

This parameterization not only enables the generative map $G_{\bm{\theta}}$ to adapt to different configurations of the McKean–Vlasov process but also provides a flexible framework to incorporate physical properties or external parameters. 

\section{Numerical experiment}\label{sec:4}
In this section, we apply the II-WGS and PI-WGS methods to five distinct examples to demonstrate their performance and compare their characteristics. The systems under consideration increase in complexity and dimensionality, and most feature mean-field interactions. 
The examples include:
a toy example, a meta-stable McKean-Vlasov systems, a mean-field model of active particles with parametric interactions, a non-gradient high-dimensional system with Gauss--Morse interaction and a high-dimensional system with Coulombic interaction.
 
In all examples, the generative map $G_{\bm{\theta}}$ is parameterized using the Real NVP architecture described in Section~\ref{sec:3.3}. Each affine coupling layer employs a fully connected neural network with three hidden layers and the LeakyReLU activation function to represent the translation and scaling functions. Across all examples, we utilize the Real NVP architecture with six affine coupling layers. Unless stated otherwise, the base distribution $\rho_{\text{base}}(\mathbf{z})$ is the standard multivariate Gaussian $\mathcal{N}(\mathbf{z};\mathbf{0},\mathbf{I}_d)$.
The accuracy of the learned stationary distribution $p_{\bm{\theta}}(\mathbf{x})$, obtained by pushing forward $\rho_{\text{base}}(\mathbf{z})$ through $G_{\bm{\theta}}$, is quantified by the relative $L^2$ error against the ground-truth distribution $p(\mathbf{x})$:
$$
e_p=\frac{\|p_{\bm{\theta}}(\mathbf{x})-p(\mathbf{x})\|_2}{\|p(\mathbf{x})\|_2}.
$$
Specific hyperparameters for the training process, such as the number of iterations $N_I$, batch size $N$, number of test functions $N_\varphi$, and test function kernel width $\kappa$, are detailed for each example.

\subsection{Example 1: Linear drift and quadratic interaction potential}\label{sec:4.1}
In this section, we consider that the dynamical system is
\begin{equation}\label{eqn:e1}
    \begin{aligned}
    \d X&=-(X-1)\dt-\nabla_{x} W *\Bar{p}_t(X,Y)\dt + \sqrt{2}\d B_1,\\
    \d Y&=-(Y-1)\dt-\nabla_{y} W *\Bar{p}_t(X,Y)\dt+  \sqrt{2}\d B_2,
    \end{aligned}
\end{equation}
where $W(\mathbf{v})=\|\mathbf{v}\|^2/2$.
This system   admits a unique stationary solution as the Gaussian distribution:
$
p(x,y)=\frac{1}{\pi}\exp\{-(x-1)^2-(y-1)^2\}
$, 
where the   mean vector $\bm{\mu}=(1,1)^\top$ and covariance matrix $\mathbf{C}=\frac{1}{2} \mathbf{I}_2$. This analytical solution serves as the ground truth for our numerical tests.

The models were trained for $N_I=50,000$ iterations with $N=10,000$ sample data points and $N_\varphi=100$ test functions. The kernel width was fixed at $\kappa=1.0$, and the learning rate decayed exponentially from an initial value of $0.001$.

\begin{figure}[htbp]
    \centering
    \includegraphics[width=0.8\textwidth]{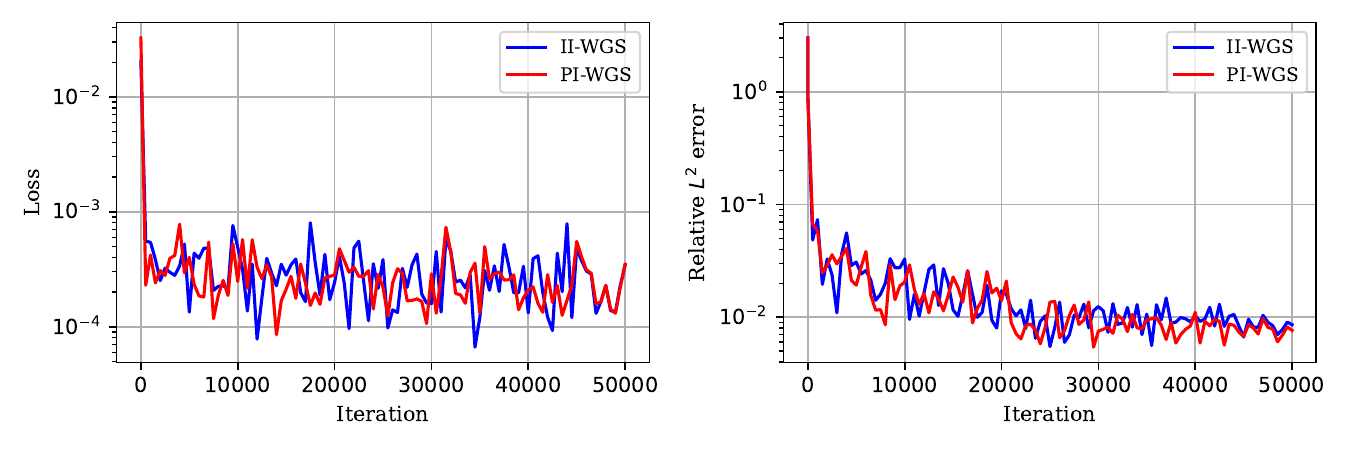}
    \caption{(Example 1) Convergence of II-WGS and PI-WGS for the linear system. Left: Training loss computed using \eqref{eqn:iiloss} and \eqref{eqn:piloss}
    versus iteration for II-WGS and PI-WGS. Right: Relative $L^2$ error versus the iteration for II-WGS and PI-WGS.}
    \label{fig:e1_loss}
\end{figure}

\begin{figure*}[htbp]
    \centering
    \includegraphics[width=0.8\textwidth]{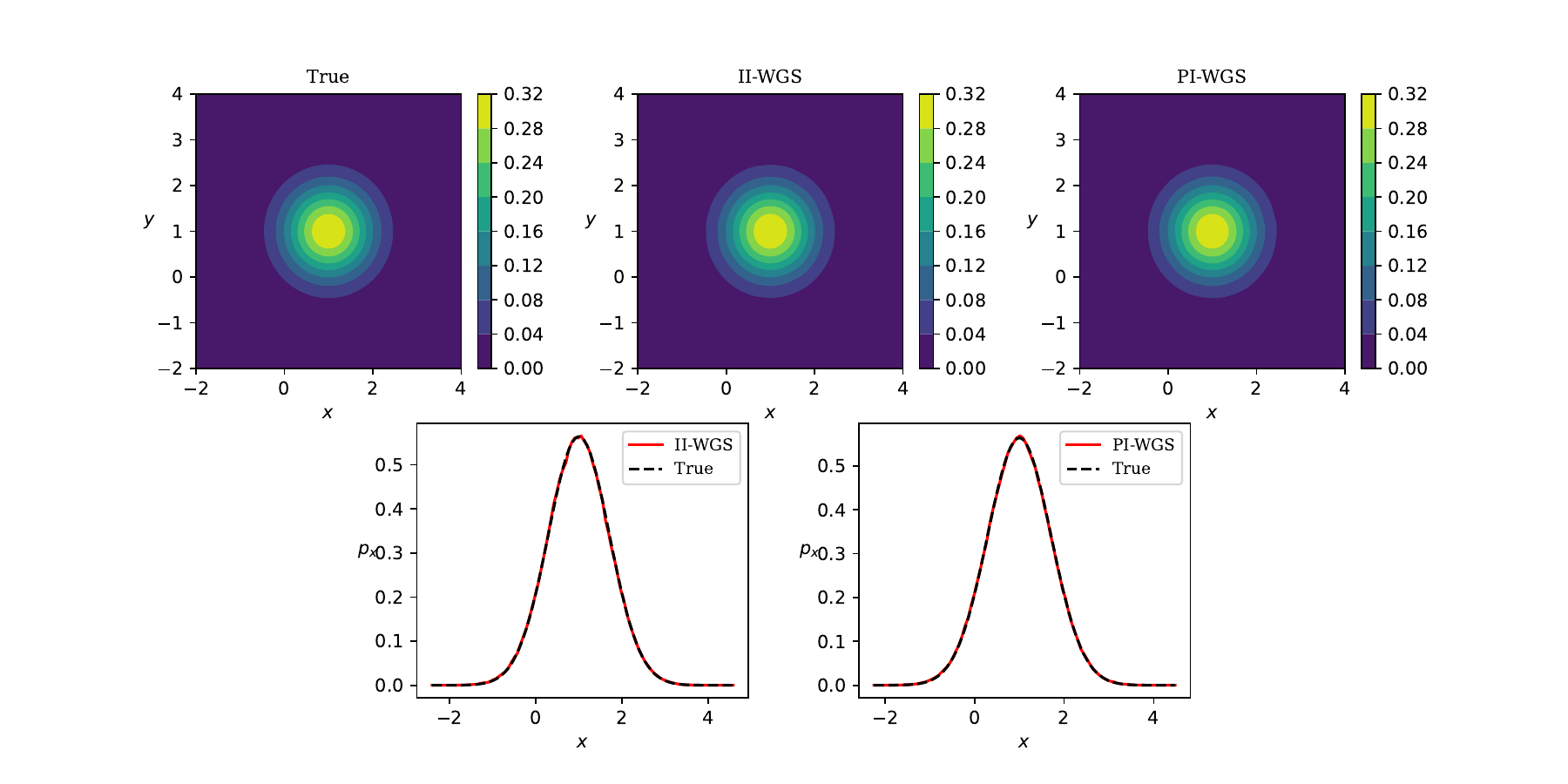}
    \caption{(Example 1) 
    Contour plots of the stationary distribution   are shown in the upper panels: ground truth distribution (upper left), II-WGS approximation (upper middle), and PI-WGS approximation (upper right). 
    The lower panels compare the computed marginal distributions in the $x$-direction with the ground truth.}
    \label{fig:e1_pdf}
\end{figure*}

As shown in Figure~\ref{fig:e1_loss}, both II-WGS and PI-WGS exhibit nearly identical convergence behavior, achieving a relative $L^2$ error of approximately $10^{-3}$ within $20000$ iterations. Figure~\ref{fig:e1_pdf} further confirms the high accuracy, showing excellent agreement between the computed densities and the analytical solution. The computed marginal distributions (lower panels) completely overlap with the ground truth, confirming the high accuracy of both methods. This result demonstrates that both schemes are equally effective for this linear problem.

\subsection{Example 2: Desai-Zwanzig Model and Phase Transitions}\label{sec:4.2}
Next, we study a prototypical system for mean-field phase transitions: a 2D extension of the classical Desai--Zwanzig model \cite{DesaiZwanzig1978}. The system is governed by a confining double-well potential $V(x, y) = (x^2 - 1)^2 + y^2$ combined with a quadratic interaction kernel $W(\mathbf{v}) = \frac{\vartheta}{2}\|\mathbf{v}\|^2$. 
The dynamics are described by the following system of stochastic differential equations  \cite{monmarche2025long}:
\begin{equation}\label{eqn:e2}
    \begin{aligned}
    \mathrm{d}X &= -\nabla_x V(X, Y) \, \mathrm{d}t - \nabla_x W * \bar{p}_t(X, Y) \, \mathrm{d}t + \sqrt{2} \, \mathrm{d}B_1, \\
    \mathrm{d}Y &= -\nabla_y V(X, Y) \, \mathrm{d}t - \nabla_y W * \bar{p}_t(X, Y) \, \mathrm{d}t + \sqrt{2} \, \mathrm{d}B_2,
    \end{aligned}
\end{equation}

Since the interaction force is effectively only dependent on the mean value, 
the ground truth of the stationary  distributions of this system can 
be precisely computed  via a fixed-point iteration method for the mean of the unknown distribution; See  Appendix~\ref{append:1}. 

\begin{figure}[htbp]
    \centering
    \includegraphics[width=0.8\textwidth]{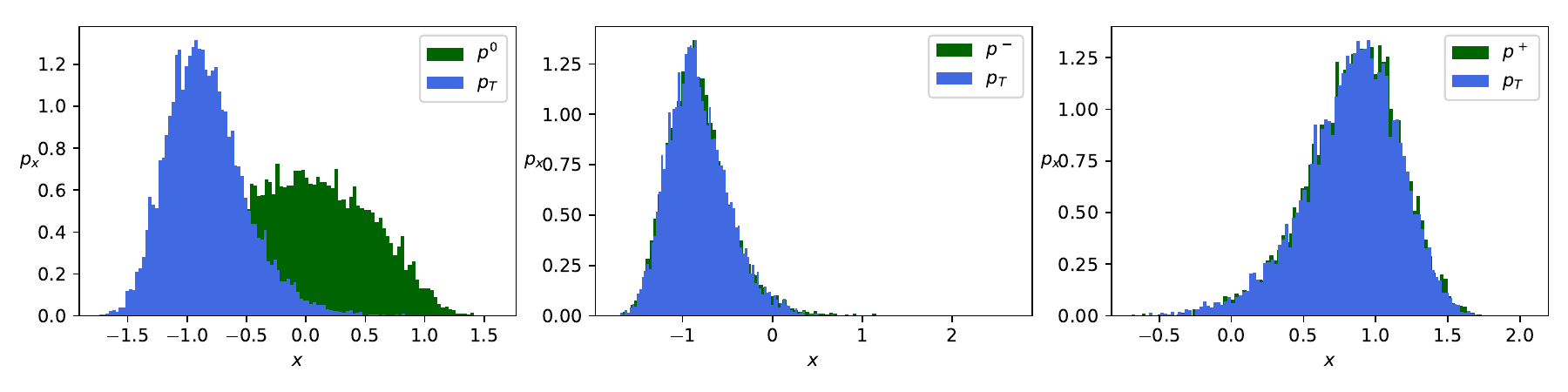}
    \caption{(Example 2 with $\vartheta = 5$): Comparison of the marginal distributions along the $x$-direction.
   The green bars represent the histograms constructed from $N=10^4$ data points sampled from the initial
  distributions $p^0$, $p^-$, and $p^+$, corresponding to the left, middle, and right panels, respectively.
  The blue bars $p_T$ represent the histogram obtained from the Euler--Maruyama simulation up to time $T=10^5$ with the time step $\delta_t=0.001$.  The results clearly show that the stationary distribution $p^0$ is unstable,
  while $p^-$ and $p^+$ are stable.}
\label{fig:e2_EM}
\end{figure}

The interaction strength $\vartheta$ plays a crucial role in determining the number of stationary distributions of the system. We select two representative values for $\vartheta$: 
\begin{itemize}
    \item  $\vartheta = 1$:  the system exhibits a \emph{unique} stationary distribution.
    \item   $\vartheta = 5$:  the system exhibits \emph{three} distinct stationary distributions $p^0$, $p^-$ and $p^+$. The stationary distribution $p^0$ is unstable, as confirmed by direct simulation (See Figure~\ref{fig:e2_EM}).
\end{itemize}

\begin{figure}
\includegraphics[width=0.9\textwidth]{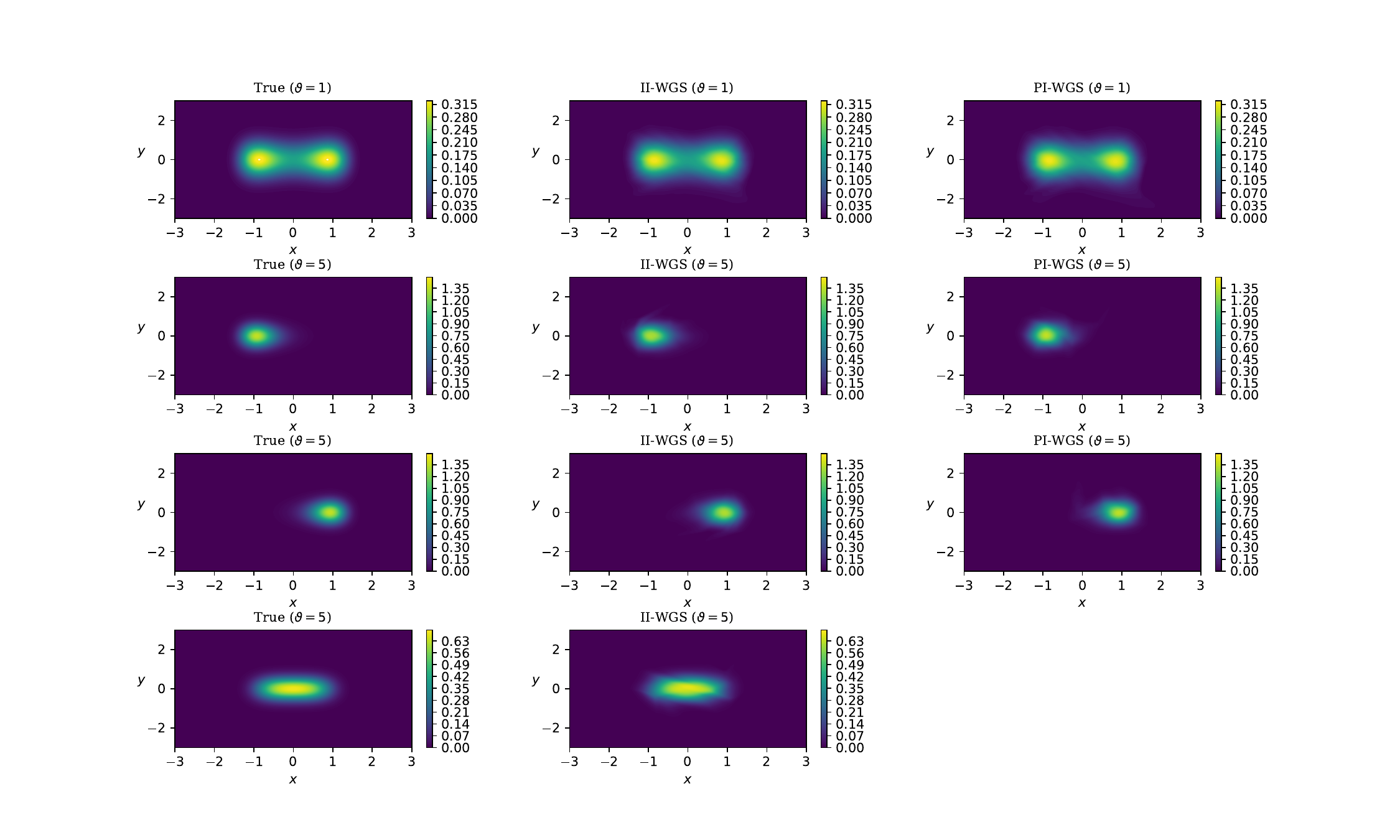}
\caption{(Example 2) Comparison of the true stationary distributions for $\vartheta = 1$ (the first row) and $\vartheta = 5$ (the other rows) with the results obtained using II-WGS and PI-WGS methods.}
\label{fig:e2_all}
\end{figure}

For both cases, training was conducted for $N_I = 50000$ iterations with $N=10000$ sample data points and $N_\varphi=100$ test functions. The learning rate is initialized at $0.001$ with exponential decay schedule during training. The hyperparameter $\kappa$ was fixed at $0.3$ for $\vartheta=1$ and $0.5$ for $\vartheta=5$.

As illustrated in Figure~\ref{fig:e2_all}, both the II-WGS and PI-WGS methods successfully converge to the unique stationary distribution in the case of $\vartheta = 1$. 
For $\vartheta = 5$, however, their behaviors diverge. 
II-WGS can converge to any of the three stationary solutions depending on initialization (like random seed), including the unstable one.  
In contrast, the PI-WGS method does not recover the unstable distribution $p^0$ in our experiments. This behavior is consistent with the local instability of the associated idealized Picard fixed-point map, as discussed in  Appendix~\ref{append:1}.

\subsection{Example 3: Active particle with mean-field interaction}\label{sec:4.3}
This example involves a mean-field model for an active particle \cite{gomes2020mean},
where $X$ is the position and $\eta$ is an active force modeled as colored noise. The state is $Z=(X, \eta)^\top$, and the dynamics is :
\begin{equation}\label{eqn:e3}
\begin{aligned}
\d X  &= \left(-X + \epsilon^{-1} \eta - \int_{\mathbb{R}}(X - X') p(X', t)  \d X' \right) \d t, \\
\d\eta &= \left(-\epsilon^{-2} \eta \right) \dt + \sqrt{2 \epsilon^{-2}} \d B_t.
\end{aligned}
\end{equation}
$B_t$ is a standard one-dimensional Brownian motion.
The Ornstein-Uhlenbeck (OU) process $\eta$ is used to model the colored noise with the stationary distribution $\mathcal{N}(0,1)$.

$\epsilon>0$ is an important parameter to characterize  the relaxation time of the active noise  and the time scale between the dynamics between $X$ and $\eta$. As $\epsilon$ tends to zero,  $\eta$ reaches the equilibrium  Gaussian distribution with zero mean and unit variance, thus  effectively the dynamics of $X$ in \eqref{eqn:e3} 
becomes the following It\^o SDE driven by the white noise:  $\d {\bar{X}} = \left( -\bar{X}   - \int (\bar{X} - \bar{X}') p(\bar{X}', t) d\bar{X}'\right)\d t  + \sqrt{2}\d B_t  $, i.e.,  $\epsilon^{-1}\eta$ is replaced by  the white noise. Indeed, for any finite $\epsilon$,  the (stationary) correlation $\mathbb{E}[\epsilon^{-1}\eta^\epsilon_{s+t}\cdot \epsilon^{-1}\eta^\epsilon_s]= \epsilon^{-2}e^{-t/\epsilon^2}$  converges to the delta function $\delta(t)$ as  $\epsilon \to 0$, indicating at a small $\epsilon$, the color noise 
$\epsilon^{-1}\eta^\epsilon_{t}$ for the first equation becomes the white noise $\d B_t$.

In fact, the stationary distribution  of the original system \eqref{eqn:e3}  can be easily computed for any $\epsilon$, which is a zero-mean Gaussian with  the following  covariance matrix $\Sigma_\epsilon$: 
\begin{small}
$$
\begin{aligned}
\Sigma_\epsilon
&=\begin{bmatrix}
\frac{1}{2(1+2\epsilon^2)} & \frac{\epsilon}{1+2\epsilon^2} \\ 
\frac{\epsilon}{1+2\epsilon^2} & 1 
\end{bmatrix}.
\end{aligned}
$$
\end{small}
which converges to 
$\begin{bmatrix}
\frac{1}{2} & 0\\
0 &  1\\
\end{bmatrix}
$
as $\varepsilon\to 0$.

Therefore, we are interested in computing the true stationary distribution of the original \eqref{eqn:e3} when $\epsilon$ is small. In particular, we treat $\epsilon$ as the additional parameter and select a range of $\epsilon$ in training our generative model while  test the performance of our method when extrapolating to even smaller value of $\epsilon$.
  Specifically,  we treat $\epsilon$ as a parameter in a \textit{parametric normalizing flow} (Section \ref{ssec332}).
The models were trained for $N_I=10,000$ iterations ($N=10,000$, $N_\varphi=100$) with parameter values sampled from $\epsilon \sim \text{Unif}[0.3, 0.6]$ and evaluated on the larger interval $[0.1, 0.8]$. The  training loss function is defined by the expected loss:
\[
\mathbb{E}_{\epsilon \sim \text{Unif}[0.3, 0.6]}\left[L\left(G_{\bm{\theta}}(\epsilon)\right)\right],
\]
where $L(G_{\bm{\theta}}(\epsilon))$ is instantiated as $L_I(G_{\bm{\theta}}(\epsilon))$ for II-WGS and as $L_P(G_{\bm{\theta}}(\epsilon))$ for PI-WGS. The parameter \( \kappa \) is initialized at $1.0$ and decays exponentially to $0.8$. The learning rate is initialized at \( 0.0001 \) and also decays exponentially during training.

\begin{figure*}
\centering
\includegraphics[width=0.88\textwidth]{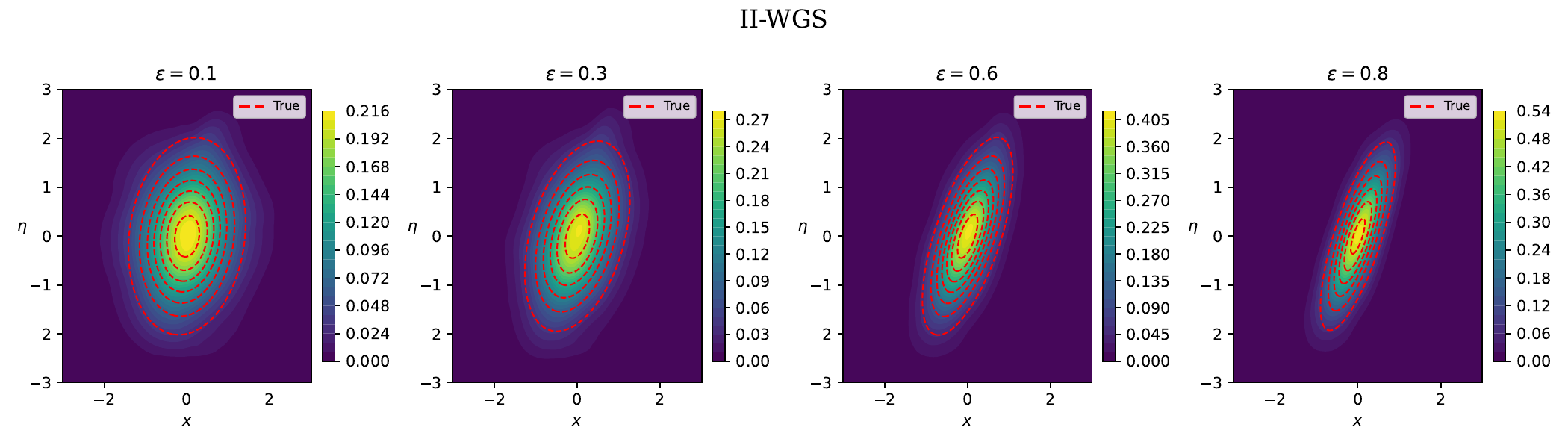}
\includegraphics[width=0.88\textwidth]{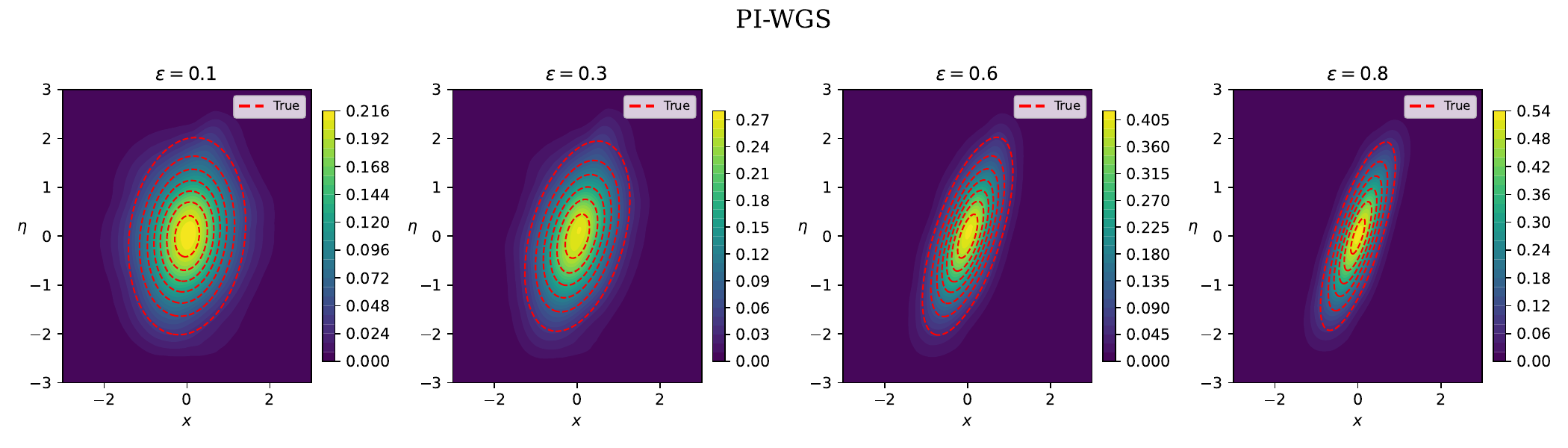}
\caption{(Example 3) Comparison of the true stationary distributions for 
$\epsilon = 0.1$, $0.3$, $0.6$, and $0.8$ with the results obtained using the II-WGS and PI-WGS methods. The contour plots represent the estimated densities obtained by II-WGS and PI-WGS, respectively, while the dashed lines indicate the true stationary densities.}
\label{fig:e3_density}
\end{figure*}

\begin{figure}
\centering
\includegraphics[width=0.8\textwidth]{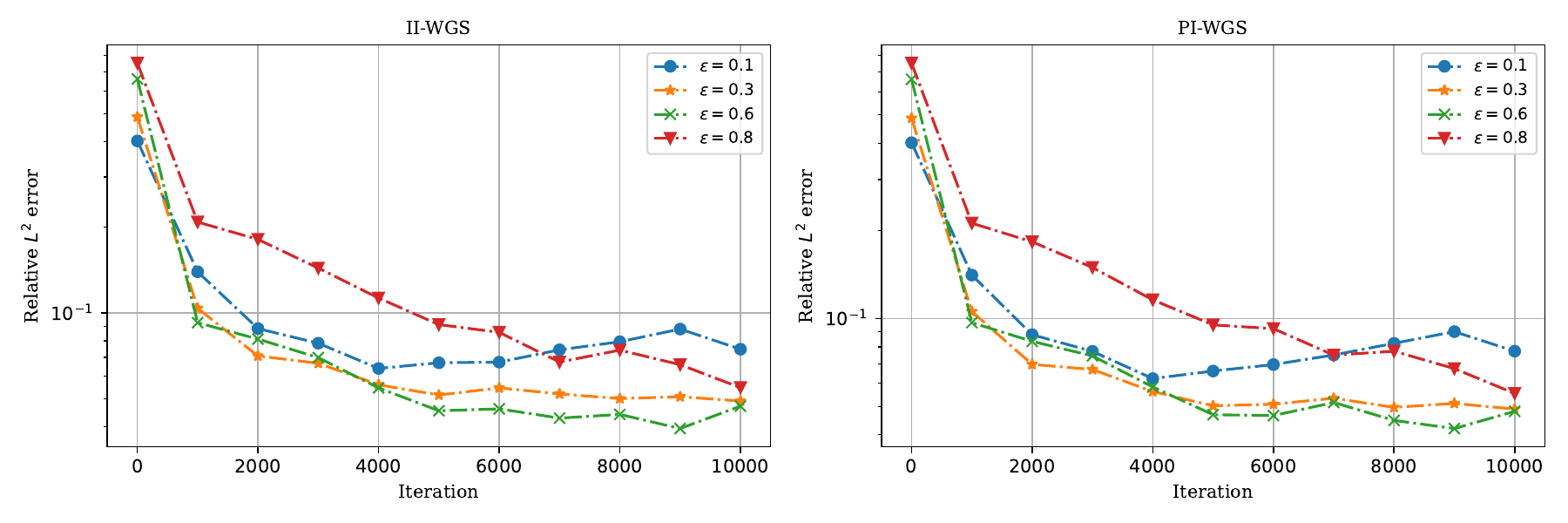}
\caption{(Example 3) Relative $L^2$ error of the learned distribution $p_{\bm{\theta}}(x)$ computed by II-WGS and PI-WGS for $\epsilon=0.1$, $0.3$, $0.6$ and $0.8$.}
\label{fig:e3_error}
\end{figure}

Figure~\ref{fig:e3_density} compares the true stationary distributions for various values of $\epsilon$ ($\epsilon = 0.1$, $0.3$, $0.6$, and $0.8$) with the estimated results obtained using the II-WGS and PI-WGS methods. The contour plots represent the estimated probability densities for both methods, while the dashed lines denote the true stationary densities. Both II-WGS and PI-WGS demonstrate the ability to approximate the true distributions across different levels of $\epsilon$. Notably, both methods accurately approximate the true distributions for various $\epsilon$.

Figure~\ref{fig:e3_error} confirms this quantitatively, with relative $L^2$ errors on the order of $10^{-2}$ for both schemes across all tested $\epsilon$ values.
 For the II-WGS method, the relative \( L^2 \) errors corresponding to \(\epsilon = 0.1\), \(0.3\), \(0.6\), and \(0.8\) are \(0.0789\), \(0.0498\), \(0.0483\), and \(0.0546\), respectively. Similarly, for the PI-WGS method, the relative \( L^2 \) errors are \(0.0807\), \(0.0479\), \(0.0487\), and \(0.0567\), respectively.

It is remarkable to see that for  $\epsilon=0.1, 0.8$, which are 
 {\it outside the training range}, the generated samples match the theoretic prediction very well,  demonstrating reasonable extrapolation performance generalizations of our method.

\subsection{Example 4: High-Dimensional Non-Gradient System with Gauss--Morse Interaction}\label{sec:4.4}
\begin{figure}
    \centering
\includegraphics[width=0.75\linewidth]{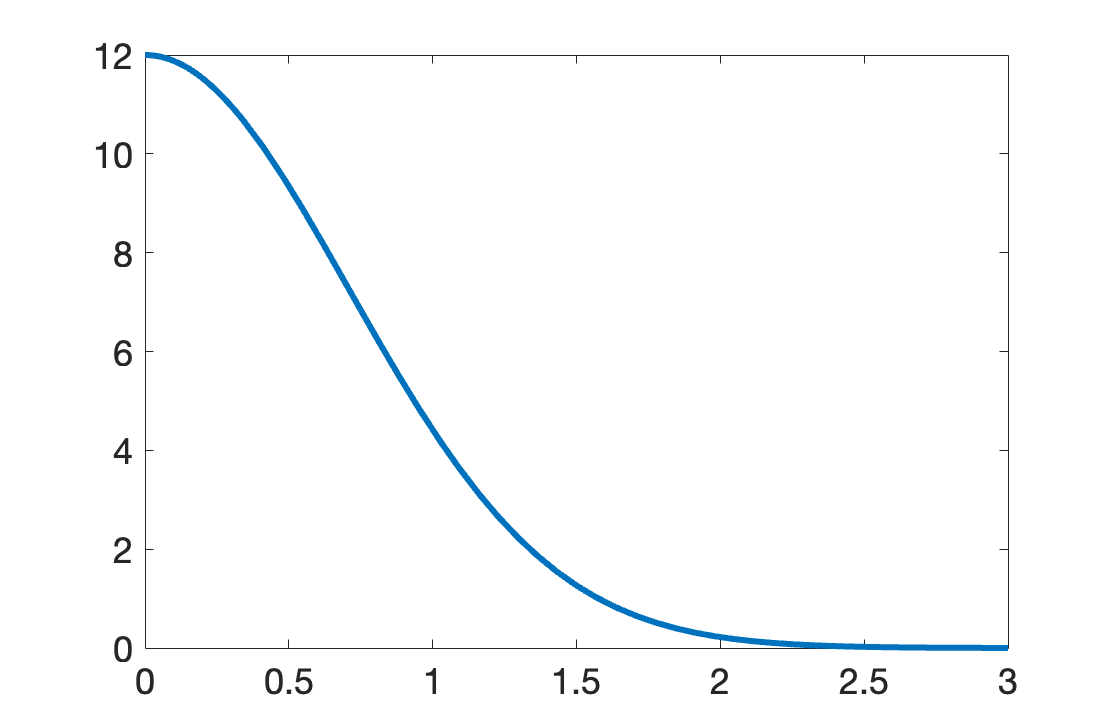}
    \caption{The example of Gauss-Morse interaction potential $W$.}
    \label{fig:GMP}
\end{figure}

We now test the methods on a 10-dimensional non-gradient system, extending the model from \cite{lin2022computing} by introducing a mean-field interaction term. The dynamics are:
\begin{equation}
\label{eqn:e4_refined}
\mathrm{d} \mathbf{X} = f(\mathbf{X}) \,\mathrm{d}t - \nabla_{\mathbf{x}}W*\bar{p}_t(\mathbf{X}) \d t+ \sqrt{2\epsilon} \mathbf{Q} \,\mathrm{d}\mathbf{B}_t,
\end{equation}
where \( \mathbf{x} \in \mathbb{R}^{10} \), \( f(\mathbf{x}) \) is the drift term (explicitly defined below). The term 
$-\nabla_{\mathbf{x}} W * \bar{p}_t$ is the newly added mean-field interaction, with
\begin{equation}
W(\mathbf{v}) = \alpha \left( C_r\, e^{-\|\mathbf{v}\|^2/\ell_r^2} 
- C_a\, e^{-\|\mathbf{v}\|^2/\ell_a^2} \right)
\end{equation}
a Gauss--Morse potential including  both repulsion and attraction parts. We set $\alpha = 10$, $C_r = 2$, $C_a = 0.8$, $\ell_r = 1$, $\ell_a = 1$. See Figure \ref{fig:GMP} for the plot of $W$. \( \mathbf{B}_t \) is a \( 10 \)-dimensional Brownian motion and  \( \epsilon = 0.1 \).

The system in~ \cite{lin2022computing} couples five identical and independent two-dimensional subsystems. 
Specifically, the dynamics for the original state \( \mathbf{Y} = (Y_1, \ldots, Y_{10})^T \) are given by \( \mathrm{d}\mathbf{Y} = g(\mathbf{Y})\,\mathrm{d}t + \sqrt{2\epsilon}\,\mathrm{d}\mathbf{B}_t \), where the drift term \( g(\mathbf{Y}) \) for the \( i \)-th subsystem (\( 1 \leq i \leq 5 \)) is defined as:
\begin{equation*}
\left\{
\begin{aligned}
&g_{2i-1}(\mathbf{Y}) = -Y_{2i-1} + Y_{2i}(1 + \sin Y_{2i-1}), \\
&g_{2i}(\mathbf{Y}) = -Y_{2i} - Y_{2i-1}(1 + \sin Y_{2i-1}).
\end{aligned}
\right.
\end{equation*}

By applying the transformation \( \hat{\mathbf{x}} = \mathbf{Q}\mathbf{y} \in \mathbb{R}^{10} \), where \( \mathbf{Q} \in \mathbb{R}^{10 \times 10} \) is a given matrix and \( \mathbf{y} = (y_1, \ldots, y_{10})^T \), the dynamics of \( \hat{\mathbf{X}} \) are governed by the following SDE:
\begin{equation}
\label{eqn:e4_hat_refined}
\mathrm{d} \hat{\mathbf{X}} = f(\hat{\mathbf{X}})\,\mathrm{d}t + \sqrt{2\epsilon} \mathbf{Q}\,\mathrm{d}\mathbf{B}_t,
\end{equation}
where the drift term \( f \) is  given by the relation \( f(\hat{\mathbf{x}}) = \mathbf{Q}g(\mathbf{Q}^{-1}\hat{\mathbf{x}}) \). The matrix \( \mathbf{Q} = [q_{ij}] \) is defined as:
\[
q_{ij} = 
\begin{cases}
0.8, & \text{if } i = j = 2k-1, \quad 1 \leq k \leq 5, \\
1.25, & \text{if } i = j = 2k, \quad 1 \leq k \leq 5, \\
-0.5, & \text{if } j = i+1, \quad 1 \leq i \leq 9, \\
0, & \text{otherwise}.
\end{cases}
\]

The system described by \eqref{eqn:e4_refined} extends the dynamics in \eqref{eqn:e4_hat_refined} by adding a mean-field interaction term.  The drift term \( f(\mathbf{x}) \) remains consistent between \eqref{eqn:e4_refined} and \eqref{eqn:e4_hat_refined}, while the noise term is scaled by \( \sqrt{2\epsilon} \).

To validate the accuracy of our methods, we estimate the stationary distribution's probability density function \( p \) using 
the Euler--Maruyama method. Specifically, we initialize the system with \( 1000 \) particles uniformly sampled from the cube \( [-2, 2]^{10} \). 
Each trajectory is simulated for a sufficiently long time horizon \( T = 10^5 \), with a fixed time step \( \delta t = 0.001 \). To ensure the system has reached a stationary regime, we discard the initial transient phase by setting a burn-in time \( T_0 = 100 \). After this time, we record the particle positions every \( 1000 \) iterations to form the empirical distribution used in the estimation of \( p \).
Training was run for $N_I=10000$ iterations ($N=10000$, $N_\varphi=100$), with a learning rate decaying from $0.0002$ and a kernel width $\kappa$ decaying  from $1.0$ to $0.9$. 

\begin{figure*}
\centering
\includegraphics[width=0.85\linewidth]{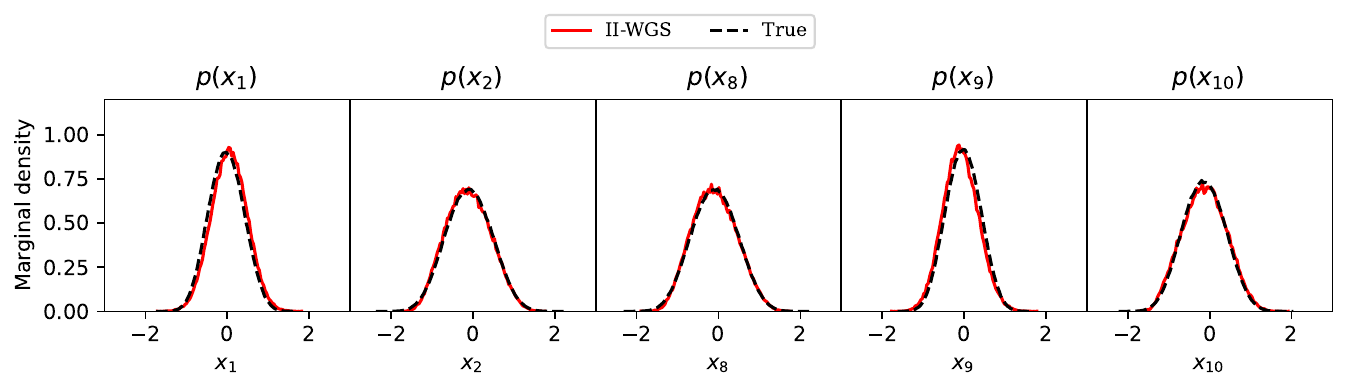}
\includegraphics[width=0.85\linewidth]{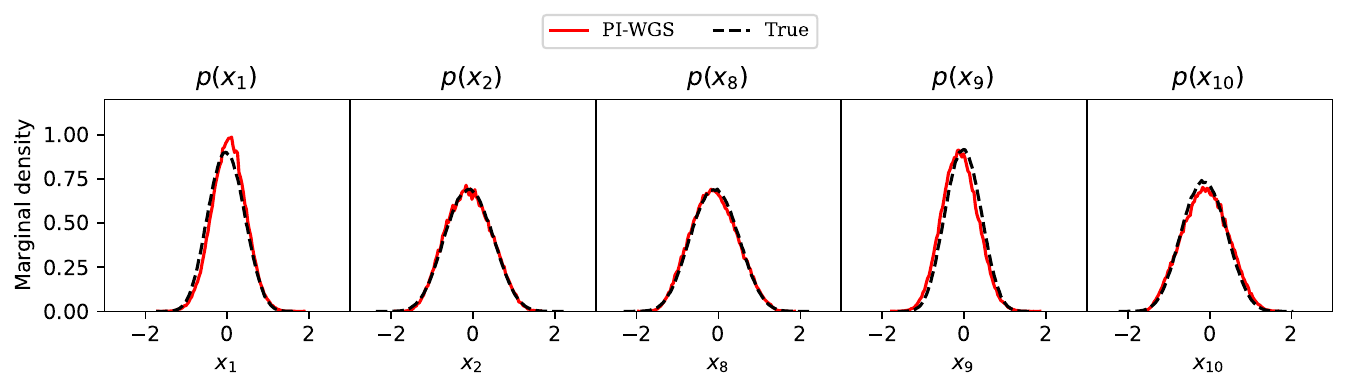}
\caption{ (Example 4) The upper and lower panels show histogram plots of the sample data points generated by the II-WGS (upper) and PI-WGS (lower) methods, respectively, compared with the true distributions obtained from SDE simulations in selected dimensions.}
\label{fig:e4}
\end{figure*}

Figure~\ref{fig:e4}  compares the marginal histograms of sample points generated by II-WGS (upper panel) and PI-WGS (lower panel) with the true stationary distributions obtained from SDE simulations. Both methods demonstrate excellent agreement with the true distributions, showcasing their comparable accuracy. This example highlights the capability of II-WGS and PI-WGS to effectively handle non-gradient systems.

\subsection{Example 5: High-Dimensional System with truncated Coulombic interaction potential}\label{sec:4.5}
Our final example is a 30-dimensional system with linear drift and a long-range, truncated Coulombic interaction:
\begin{equation}
\label{eqn:e5}
\begin{aligned}
\mathrm{d} \mathbf{X} = -(\mathbf{X} - \boldsymbol{\mu})\, \mathrm{d}t - \left( K * \bar{p}_t \right)(\mathbf{X})\, \mathrm{d}t + \sqrt{2}\bm{\Sigma}^{\frac{1}{2}}\, \mathrm{d}\mathbf{B}_t,
\end{aligned}
\end{equation}
where \(\mathbf{x} \in \mathbb{R}^{30}\), \(\mathbf{B}_t\) denotes a standard 30-dimensional Brownian motion, and the interaction kernel \(K(\mathbf{x}, \mathbf{y})\) is defined as the truncated Coulombic interaction potential \( K(\mathbf{x}, \mathbf{y}) \):
\[
K(\mathbf{x}, \mathbf{y}) = \alpha \frac{\mathbf{x} - \mathbf{y}}{c + \|\mathbf{x} - \mathbf{y}\|^d},
\]
where \( c >0 \) is a small constant to ensure that \( K(\mathbf{x}, \mathbf{y}) \) remains bounded as \( \mathbf{x} = \mathbf{y} \), \( \alpha > 0 \) controls the interaction strength, and \( d > 0 \) determines the decay rate of the interaction. In this example, we set $c=10^{-6}$, $\alpha=0.2$ and $d=30$.

Here, \(\boldsymbol{\mu} \in \mathbb{R}^{30}\) is a vector whose entries \(\mu_i\) are randomly drawn from the uniform distribution \(\text{Unif}[-1, 1]\) and then is fixed. The  matrix \(\bm{\Sigma} \in \mathbb{R}^{30 \times 30}\) is diagonal, with each diagonal entry defined as $1/i$ for $i=1,\cdots,30$.

A reference solution was generated using the same particle-based simulation strategy described in Section~\ref{sec:4.4}. Specifically, the system was initialized with \( 1000 \) particles uniformly sampled from the cube \( [-1, 1]^{30} \). The subsequent evolution followed the same procedure as outlined in Section~\ref{sec:4.4}.

The training process is performed over \( N_I = 10000 \) iterations, using \( N = 10000 \) sample points and \( N_\varphi = 100 \) test functions. Throughout the training phase, we use two groups of $\kappa$ in test functions. The first group is
characterized by a fixed $\kappa = 10$, while the second type starts with the same value but follows an exponential decay schedule. $\gamma$ was initially set at $0.8$ and then gradually decreased to $0.08$ during the training process. Furthermore, the learning rate was set to $0.0001$ and followed an exponential decay schedule.

\begin{figure*}
\centering
\includegraphics[width=0.85\linewidth]{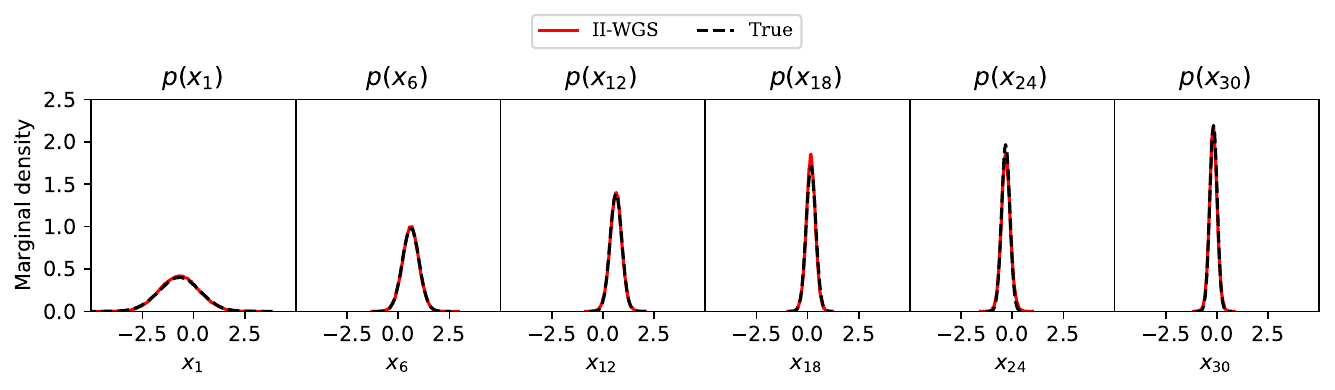}
\includegraphics[width=0.85\linewidth]{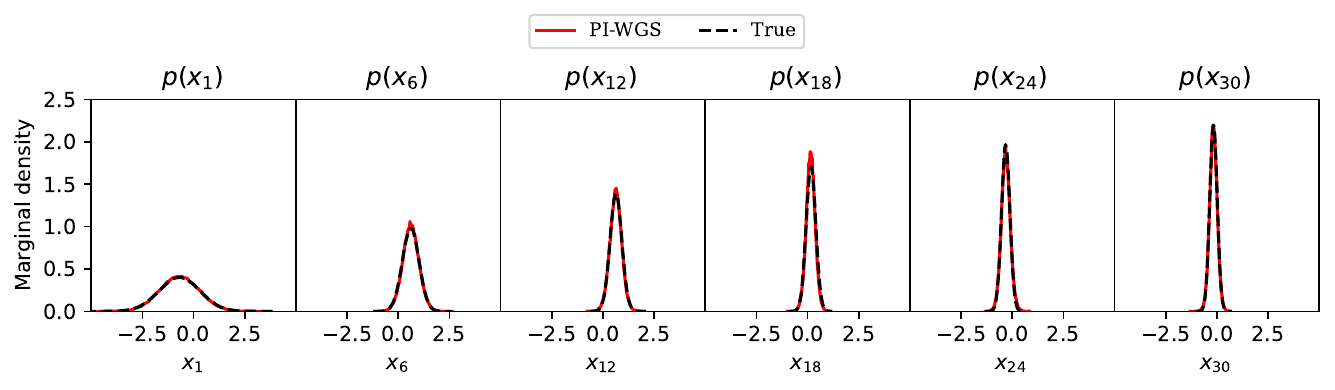}
\caption{ (Example 5) The upper and lower panels show histogram plots of the sample data points generated by the II-WGS (upper) and PI-WGS (lower) methods, respectively, compared with the true distributions obtained from SDE simulations in selected dimensions.}
\label{fig:e5}
\end{figure*}

\begin{figure*}
\centering
\begin{minipage}{0.49\linewidth}
    \centering
    \includegraphics[width=\linewidth]{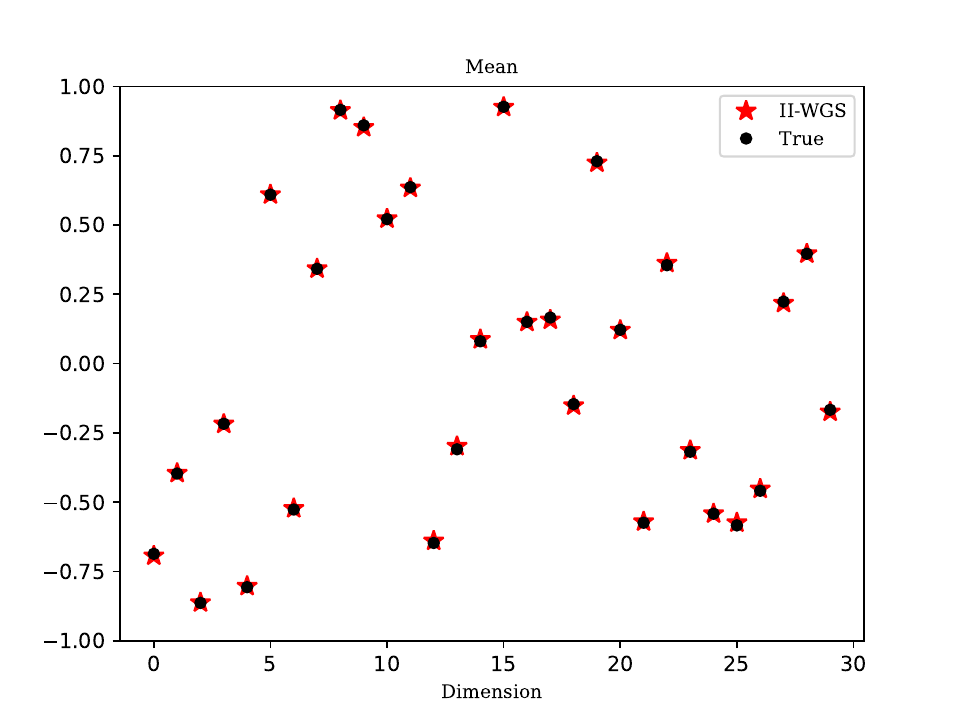}
\end{minipage}
\hfill
\begin{minipage}{0.49\linewidth}
    \centering
    \includegraphics[width=\linewidth]{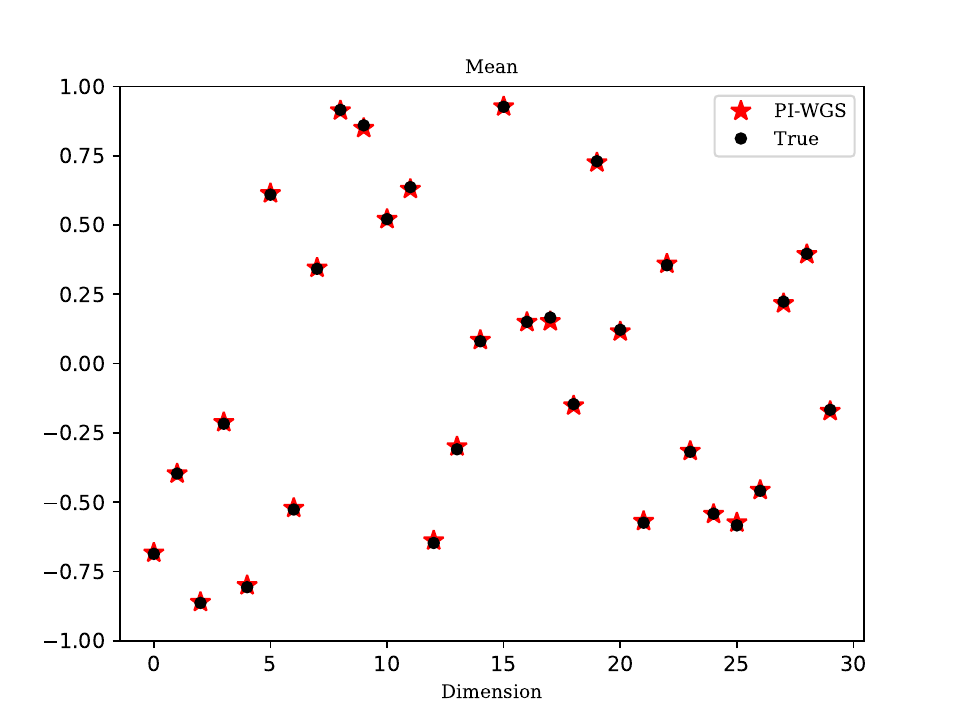}
\end{minipage}
\caption{(Example 5) The left and right panels display the estimated means (red stars) with the true means (black points) in each dimension by II-WGS (left) and PI-WGS (right), respectively.}
\label{fig:e5-mean}
\end{figure*}

Figure~\ref{fig:e5} compares the marginal histograms of sample points generated by II-WGS (upper panel) and PI-WGS (lower panel) with the true stationary distributions from SDE simulations. Both methods show excellent agreement with the true distributions, and their performances are nearly identical. Figure~\ref{fig:e5-mean} compares the estimated means (red stars) with the true means (black points) for II-WGS (left) and PI-WGS (right). The results indicate that both methods produce nearly indistinguishable mean estimates, consistent with the true values across all dimensions, confirming the methods' effectiveness for high-dimensional systems with singular, long-range interactions.

\section{Conclusions}\label{sec:5}
In this work, we demonstrate how the Weak Generative Sampler (WGS) framework  \cite{cai2026weak} effectively addresses the challenge of sampling from the stationary measure of mean-field interacting systems. For the McKean-Vlasov system, our approach constructs a generative map derived from the weak form of the nonlinear stationary Fokker-Planck equation. The proposed method successfully overcomes the intrinsic challenges of solving this equation, which include generating i.i.d. samples, handling high-dimensional state spaces, managing nonlocal interaction terms, and avoiding the errors inherent in finite-particle approximations.

When multiple stationary distributions exist, the specific distribution to which the algorithm converges is inherently dependent on the initialization of the generative neural network. This behaviour is analogous to that of any local optimization method applied to a non-convex objective function. To explore the full landscape of possible stationary measures, the simulation of finite-particle systems is often infeasible, as the resulting random empirical measures—which transition stochastically between metastable states—are difficult to analyze systematically. An alternative strategy involves injecting artificial random perturbations into the mean-field model itself. For instance,  \cite{Kolodziejczyk2025kc} employs a spectral Galerkin scheme to solve a two-dimensional stochastic nonlinear Fokker-Planck equation. A similar idea could be integrated into our framework by introducing random perturbations directly into the transport maps during training.
Other future research directions include extending the proposed methods to broader classes of mean-field systems, such as those driven by non-Gaussian noise or featuring more intricate interaction kernels.

  \section*{Acknowledgments}
Xiang ZHOU acknowledges the supported by  General Research Funds from the Research Grants Council of the Hong Kong Special Administrative Region, China (Project No. 11318522,11308323, 11304525)  

\appendix
\section{Picard fixed-point iteration for Example 2}
\label{append:1}

In this appendix, we analyze the Picard-type fixed-point iteration associated with Example~2. 
In the gradient setting considered there, stationary densities of the McKean--Vlasov equation are characterized by a self-consistent Gibbs form, which naturally induces a nonlinear fixed-point map on probability densities.

We consider the map
\[
\mathcal{T} : \mathcal{P}_{2}^{ac}(\mathbb{R}) \to \mathcal{P}_{2}^{ac}(\mathbb{R}),
\]
where
\[
\mathcal{P}_{2}^{ac}(\mathbb{R})
:=
\left\{
p \in L^1(\mathbb{R})
\;\middle|\;
p(x)\ge 0,\;
\int_{\mathbb{R}} p(x)\,\mathrm{d}x = 1,\;
\int_{\mathbb{R}} x^2 p(x)\,\mathrm{d}x < \infty
\right\}
\]
denotes the space of probability densities on \(\mathbb{R}\) with finite second moment.

For a given confining potential \(V:\mathbb{R}\to\mathbb{R}\), interaction strength \(\vartheta>0\), and noise level (temperature) \(\epsilon>0\), we define
\begin{equation}
\label{eq:T-definition}
\mathcal{T}(p)(x)
=
\frac{1}{Z_p}
\exp\left(
-\epsilon^{-1}V(x)
-\frac{\epsilon^{-1}\vartheta}{2}
\int_{\mathbb{R}} (x-y)^2 p(y)\,\mathrm{d}y
\right),
\end{equation}
where $Z_p$ ensures normalization:
\[
Z_p
=
\int_{\mathbb{R}}
\exp\left(
-\epsilon^{-1}V(x)
-\frac{\epsilon^{-1}\vartheta}{2}
\int_{\mathbb{R}} (x-y)^2 p(y)\,\mathrm{d}y
\right)\,\mathrm{d}x.
\]

In this setting, stationary densities correspond exactly to the fixed points of \(\mathcal{T}\). This motivates the Picard iteration
\[
p_{n+1}=\mathcal{T}(p_n),\qquad n\ge 0,
\]
started from some initial density \(p_0\in\mathcal{P}_{2}^{ac}(\mathbb{R})\).
A key simplification arises from the observation that \(\mathcal{T}(p)\) depends on \(p\) strictly through its first moment. Indeed, letting
\[
m_p := \int_{\mathbb{R}} y\,p(y)\,\mathrm{d}y,
\]
we can expand the interaction term as
\[
\int_{\mathbb{R}} (x-y)^2 p(y)\,\mathrm{d}y
=
x^2-2xm_p+\int_{\mathbb{R}} y^2 p(y)\,\mathrm{d}y
=
(x-m_p)^2+\mathrm{Var}_p(y),
\]
where \(\mathrm{Var}_p(y) := \int_{\mathbb{R}} y^2p(y)\,\mathrm{d}y - m_p^2\). Since \(\mathrm{Var}_p(y)\) is independent of \(x\), it factors out of the exponential and is perfectly absorbed into the normalization constant. Therefore, \eqref{eq:T-definition} can be equivalently rewritten as
\begin{equation}
\label{eq:T-mean-form}
\mathcal{T}(p)(x)
=
\frac{1}{\bar Z(m_p)}
\exp\left(
-\epsilon^{-1}V(x)
-\frac{\epsilon^{-1}\vartheta}{2}(x-m_p)^2
\right),
\end{equation}
where
\[
\bar Z(m)
=
\int_{\mathbb{R}}
\exp\left(
-\epsilon^{-1}V(x)
-\frac{\epsilon^{-1}\vartheta}{2}(x-m)^2
\right)\,\mathrm{d}x.
\]

Accordingly, for each \(m\in\mathbb{R}\), we define the parameterized density
\begin{equation}
\label{eq:rho-m}
\rho_m(x)
:=
\frac{1}{\bar Z(m)}
\exp\left(
-\epsilon^{-1}V(x)
-\frac{\epsilon^{-1}\vartheta}{2}(x-m)^2
\right),
\end{equation}
which implies \(\mathcal{T}(p)=\rho_{m_p}\). Consequently, by defining the mean sequence
\[
m_n:=\int_{\mathbb{R}} x\,p_n(x)\,\mathrm{d}x,
\]
the infinite-dimensional iteration \(p_{n+1}=\mathcal{T}(p_n)\) systematically collapses into a purely scalar iteration:
\begin{equation}
\label{eq:mean-iteration}
m_{n+1}=F(m_n),
\qquad \text{where} \quad
F(m):=\int_{\mathbb{R}} x\,\rho_m(x)\,\mathrm{d}x.
\end{equation}
The fixed points of \(\mathcal{T}\) are therefore fundamentally characterized by the scalar self-consistency equation \(m=F(m)\).
For Example~2 in Section~\ref{sec:4.2}, the confining potential is dimensionally separable:
\[
V(x,y)=V_x(x)+V_y(y),
\qquad
V_x(x)=(x^2-1)^2,
\qquad
V_y(y)=y^2.
\]
Because both the confining potential and the quadratic interaction term decompose additively across the coordinates, the corresponding self-consistent Gibbs density immediately factorizes as \(\tilde p(x,y)=\tilde p_x(x)\,\tilde p_y(y)\). Here, \(\tilde p_x\) and \(\tilde p_y\) are the one-dimensional stationary densities obtained by applying the fixed-point iteration independently in each respective coordinate.

For numerical implementation, we approximate the exact fixed-point map on \(\mathbb{R}\) by truncating the spatial domain to \([-L, L]\) and evaluating the integrals via numerical quadrature, as summarized in Algorithm~\ref{alg:fixed_point_iteration}.

\begin{algorithm}[htbp]
\footnotesize
\caption{Picard fixed-point iteration for the McKean--Vlasov equation}
\label{alg:fixed_point_iteration}
\SetKwInOut{Input}{Input}
\SetKwInOut{Output}{Output}

\Input{Initial guess \(m_0\in\mathbb{R}\); truncation interval \([-L,L]\); number of iterations \(N_I\)}
\For{\(n=1\) \KwTo \(N_I\)}{
  Evaluate the unnormalized density:
  \[
  p^{(n)}(x)
  :=
  \exp\left(
  -\epsilon^{-1}V(x)
  -\frac{\epsilon^{-1}\vartheta}{2}(x-m_{n-1})^2
  \right),
  \qquad x\in[-L,L].
  \]

  Compute the normalization constant (e.g., via the trapezoidal rule):
  \[
  \bar Z^{(n)}
  :=
  \int_{-L}^{L} p^{(n)}(x)\,\mathrm{d}x.
  \]

  Normalize the density:
  \[
  \tilde p^{(n)}(x)
  :=
  \frac{1}{\bar Z^{(n)}}\,p^{(n)}(x).
  \]

  Update the mean:
  \[
  m_n
  :=
  \int_{-L}^{L} x\,\tilde p^{(n)}(x)\,\mathrm{d}x.
  \]
}
\Output{Approximate stationary density \(\tilde p^{(N_I)}\) on \([-L,L]\).}
\end{algorithm}

We now rigorously analyze the local convergence of the scalar fixed-point iteration \eqref{eq:mean-iteration}. For the double-well potential \(V_x(x)=(x^2-1)^2\), the symmetric state \(m=0\) is unconditionally a fixed point. Indeed, since \(V_x\) is an even function, the density \(\rho_0\) is symmetric, naturally yielding \(F(0)=\int_{\mathbb{R}} x\,\rho_0(x)\,\mathrm{d}x=0\).
To study the local stability of the Picard iteration \(m_{n+1}=F(m_n)\), we examine its derivative \(F'(m)\).

\begin{proposition}
\label{prop:contractive}
Assume that \(V\) is sufficiently smooth and confining such that \(\bar Z(m)<\infty\) for all \(m\in\mathbb{R}\), and differentiation under the integral sign is justified. Then the map \(F\) defined in \eqref{eq:mean-iteration} is differentiable and satisfies
\begin{equation}
\label{eq:derivative}
F'(m)
=
\epsilon^{-1}\vartheta\,\mathrm{Var}_{\rho_m}(x).
\end{equation}
In particular, \(F'(m)\ge 0\) for all \(m\), implying that \(F\) is non-decreasing.
\end{proposition}

\begin{proof}
Let \(H_m(x) := \epsilon^{-1}V(x) + \frac{\epsilon^{-1}\vartheta}{2}(x-m)^2\), such that \(\rho_m(x)=\bar Z(m)^{-1}e^{-H_m(x)}\) and \(\bar Z(m)=\int_{\mathbb{R}} e^{-H_m(x)}\,\mathrm{d}x\).
By definition, \(F(m)=\int_{\mathbb{R}} x\,\rho_m(x)\,\mathrm{d}x\). Differentiating under the integral sign and employing the standard covariance identity for Gibbs-type measures yields
\[
F'(m)
=
\mathrm{Cov}_{\rho_m}\!\left(x,-\partial_m H_m(x)\right).
\]
Since \(-\partial_m H_m(x)=\epsilon^{-1}\vartheta(x-m)\), we readily obtain
\[
F'(m)
=
\mathrm{Cov}_{\rho_m}\!\bigl(x,\epsilon^{-1}\vartheta(x-m)\bigr)
=
\epsilon^{-1}\vartheta\,\mathrm{Cov}_{\rho_m}(x,x-m).
\]
Because the shift by a constant \(m\) does not alter the covariance with \(x\), this simplifies directly to the variance:
\[
F'(m)=\epsilon^{-1}\vartheta\,\mathrm{Var}_{\rho_m}(x),
\]
which proves \eqref{eq:derivative}.
\end{proof}

As a direct consequence, if \(m^*\) is a fixed point of \(F\), then the Picard iteration is locally attracting whenever the strict inequality
\begin{equation}
\label{eq:contractive}
\epsilon^{-1}\vartheta\,\mathrm{Var}_{\rho_{m^*}}(x)<1
\end{equation}
holds. If \(F'\) is continuous in a neighborhood of \(m^*\), condition \eqref{eq:contractive} guarantees that \(F\) is locally contractive near \(m^*\).
Crucially, condition \eqref{eq:contractive} explicitly demonstrates that the local convergence of the fixed-point iteration is fundamentally governed by the interplay between the noise level \(\epsilon\) and the interaction strength \(\vartheta\). At the symmetric fixed point \(m=0\), the stability threshold is simply determined by whether \(F'(0)=\epsilon^{-1}\vartheta\,\mathrm{Var}_{\rho_0}(x)\) is smaller or larger than \(1\).

\begin{figure*}[htbp]
\centering
\includegraphics[width=0.32\linewidth]{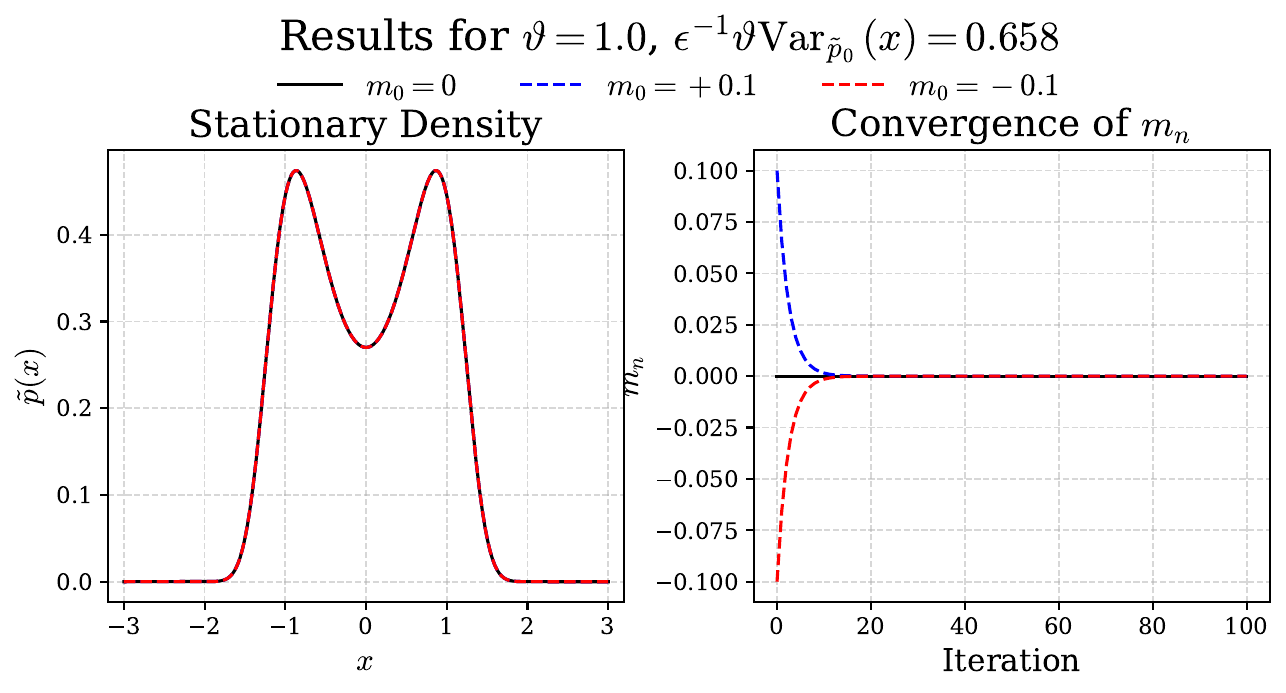}\hfill
\includegraphics[width=0.32\linewidth]{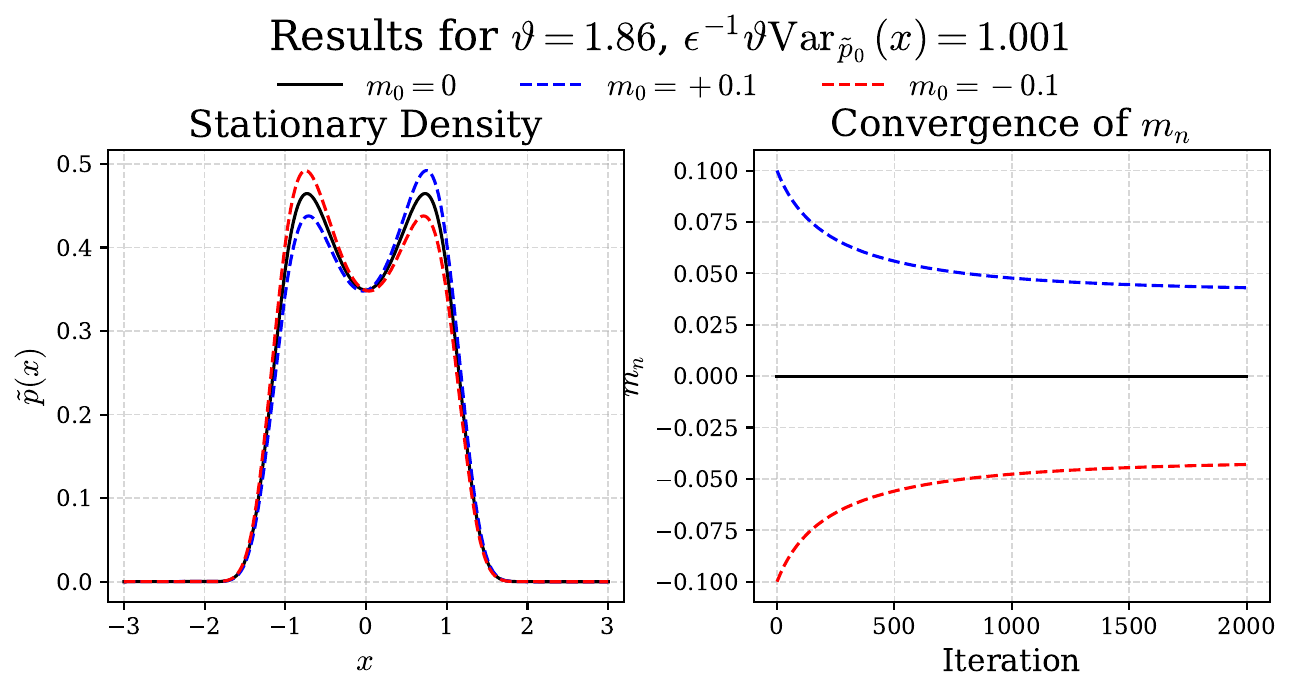}\hfill
\includegraphics[width=0.32\linewidth]{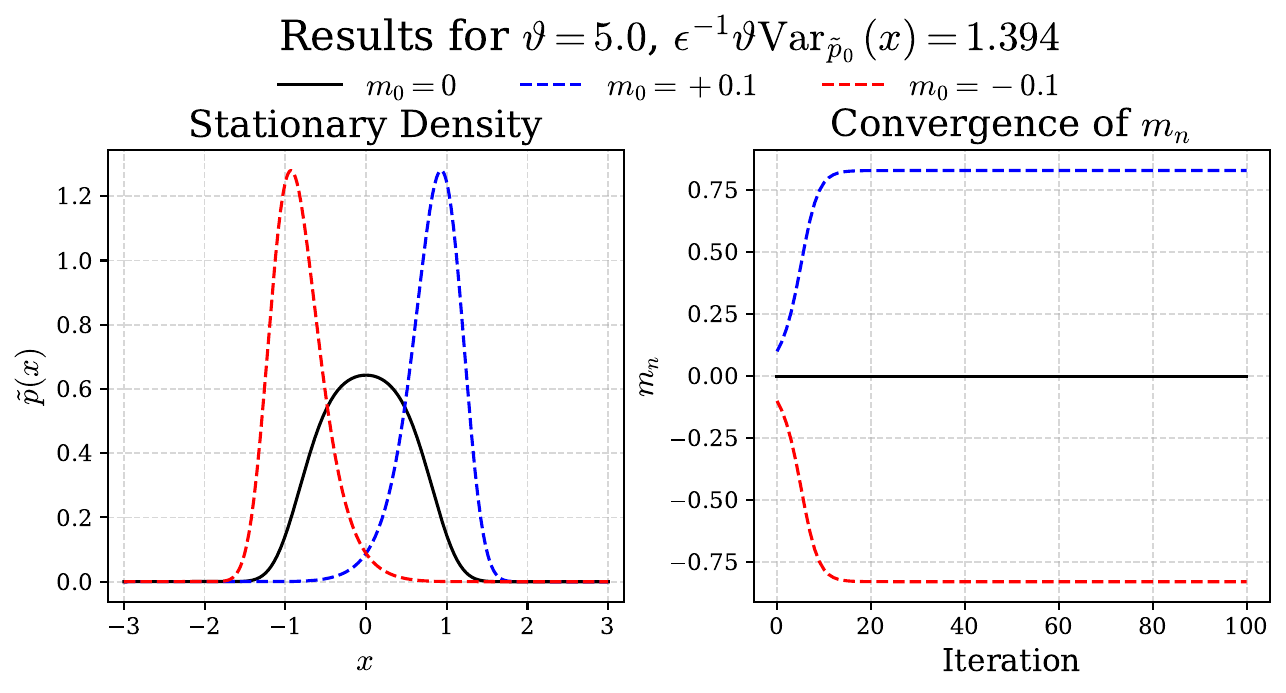}
\caption{
Stationary densities computed by Algorithm~\ref{alg:fixed_point_iteration} and the corresponding evolution of the iterates \(m_n\). Each panel represents a different interaction strength \(\vartheta\), with fixed temperature \(\epsilon=1\). The left side of each subfigure displays the final stationary density for various initializations \(m_0\), while the right side tracks the discrete Picard trajectory. The supercritical cases (\(\vartheta=1.86, 5.00\)) clearly demonstrate the loss of local attractivity for the symmetric branch.
}
\label{fig:appendix}
\end{figure*}

As illustrated in Figure~\ref{fig:appendix}, the qualitative behavior of the map \(F\) bifurcates significantly as \(\vartheta\) varies. For instance, when \(\vartheta=1\) (subcritical regime), the iteration robustly converges to the symmetric fixed point regardless of the initialization. Conversely, when \(\vartheta=1.86\) or \(\vartheta=5\) (supercritical regime), initializations near \(m=0\) inevitably diverge away and converge instead to the stable symmetry-broken fixed points, empirically confirming that the symmetric fixed point has lost its local attractivity.

This analysis explains a mechanism consistent with the observed behavior of PI-WGS in Example 2. Since PI-WGS uses a frozen-interaction update, its behavior is heuristically related to the local stability properties of the associated scalar Picard map.

\bibliography{Ref-CICP}

\begin{thebibliography}{10}

\bibitem{binney2011galactic}
James Binney and Scott Tremaine.
\newblock {\em Galactic dynamics}.
\newblock Princeton university press, 2011.

\bibitem{bittencourt2013fundamentals}
Jos{\'e}~A Bittencourt.
\newblock {\em Fundamentals of plasma physics}.
\newblock Springer Science \& Business Media, 2013.

\bibitem{boffi2023probability}
Nicholas~M Boffi and Eric Vanden-Eijnden.
\newblock Probability flow solution of the {F}okker--{P}lanck equation.
\newblock {\em Machine Learning: Science and Technology}, 4(3):035012, 2023.

\bibitem{boffi2024deep}
Nicholas~M Boffi and Eric Vanden-Eijnden.
\newblock Deep learning probability flows and entropy production rates in active matter.
\newblock {\em Proceedings of the National Academy of Sciences}, 121(25):e2318106121, 2024.

\bibitem{bolley2013uniform}
Fran{\c{c}}ois Bolley, Ivan Gentil, and Arnaud Guillin.
\newblock Uniform convergence to equilibrium for granular media.
\newblock {\em Archive for Rational Mechanics and Analysis}, 208:429--445, 2013.

\bibitem{cai2026weak}
Zhiqiang Cai, Yu~Cao, Yuanfei Huang, and Xiang Zhou.
\newblock Weak generative sampler to efficiently sample invariant distribution of stochastic differential equation.
\newblock {\em SIAM Journal on Scientific Computing}, 48(4):C708--C735, 2026.

\bibitem{carrillo2003kinetic}
Jos{\'e}~A Carrillo, Robert~J McCann, and C{\'e}dric Villani.
\newblock Kinetic equilibration rates for granular media and related equations: entropy dissipation and mass transportation estimates.
\newblock {\em Revista Matematica Iberoamericana}, 19(3):971--1018, 2003.

\bibitem{carrillo2020long}
Jose~Antonio Carrillo, Rishabh~S Gvalani, Grigorios~A Pavliotis, and Andre Schlichting.
\newblock {Long-time behaviour and phase transitions for the McKean--Vlasov equation on the torus}.
\newblock {\em Archive for Rational Mechanics and Analysis}, 235(1):635--690, 2020.

\bibitem{ChaintronReview2022I}
Louis-Pierre Chaintron and Antoine Diez.
\newblock {Propagation of chaos: A review of models, methods and applications. {I}. Models and methods}.
\newblock {\em Kinetic and Related Models}, 15(6):895--1015, 2022.

\bibitem{ChaintronReview2022II}
Louis-Pierre Chaintron and Antoine Diez.
\newblock {Propagation of chaos: A review of models, methods and applications. {II}. Applications}.
\newblock {\em Kinetic and Related Models}, 15(6):1017--1173, 2022.

\bibitem{Dawson1983}
Donald~A. Dawson.
\newblock Critical dynamics and fluctuations for a mean-field model of cooperative behavior.
\newblock {\em Journal of Statistical Physics}, 31(1):29--85, 1983.

\bibitem{DW1989}
Donald~A. Dawson and J\"{u}rgen G\"{a}rtner.
\newblock {\em Large deviations, free energy functional and quasi-potential for a mean field model of interacting diffusions}, volume~78.
\newblock Memoirs of the American Mathematical Society, 1989.

\bibitem{DesaiZwanzig1978}
Rashmi. Desai and Robert Zwanzig.
\newblock Statistical mechanics of a nonlinear stochastic model.
\newblock {\em Journal of Statistical Physics}, 19(1):1--24, 1978.

\bibitem{dinh2017density}
Laurent Dinh, Jascha Sohl-Dickstein, and Samy Bengio.
\newblock {Density estimation using Real NVP}.
\newblock In {\em International Conference on Learning Representations}, 2017.

\bibitem{duerinckx2020mean}
Mitia Duerinckx and Sylvia Serfaty.
\newblock Mean field limit for coulomb-type flows.
\newblock {\em Duke Mathematical Journal}, 169(15):2887--2935, 2020.

\bibitem{frank2005nonlinear}
T.D. Frank.
\newblock {\em {Nonlinear Fokker-Planck Equations: Fundamentals and Applications}}.
\newblock Springer Series in Synergetics. Springer Berlin Heidelberg, 2005.

\bibitem{gomes2020mean}
Susana~N Gomes, Grigorios~A Pavliotis, and Urbain Vaes.
\newblock Mean field limits for interacting diffusions with colored noise: phase transitions and spectral numerical methods.
\newblock {\em Multiscale Modeling \& Simulation}, 18(3):1343--1370, 2020.

\bibitem{guillin2024uniform}
Arnaud Guillin, Pierre Le~Bris, and Pierre Monmarch{\'e}.
\newblock {Uniform in time propagation of chaos for the 2D vortex model and other singular stochastic systems}.
\newblock {\em Journal of the European Mathematical Society}, pages 1--28, 2024.

\bibitem{guillin2022uniform}
Arnaud Guillin, Wei Liu, Liming Wu, and Chaoen Zhang.
\newblock {Uniform Poincar{\'e} and logarithmic Sobolev inequalities for mean field particle systems}.
\newblock {\em The Annals of Applied Probability}, 32(3):1590--1614, 2022.

\bibitem{GVALANI2020}
Rishabh~S. Gvalani and André Schlichting.
\newblock Barriers of the {McKean–Vlasov} energy via a mountain pass theorem in the space of probability measures.
\newblock {\em Journal of Functional Analysis}, 279(11):108720, 2020.

\bibitem{LevyEPR2026PRL}
Yuanfei Huang, Chengyu Liu, Bing Miao, and Xiang Zhou.
\newblock {Entropy Production in Non-Gaussian Active Matter: A Unified Fluctuation Theorem and Deep Learning Framework}.
\newblock {\em Phys. Rev. Lett.}, 136:068302, Feb 2026.

\bibitem{jabin2014review}
Pierre-Emmanuel Jabin.
\newblock {A review of the mean field limits for Vlasov equations}.
\newblock {\em Kinetic and Related models}, 7(4):661--711, 2014.

\bibitem{Jin2020RBM}
Shi Jin, Lei Li, and Jian-Guo Liu.
\newblock Random {Batch Methods} ({RBM}) for interacting particle systems.
\newblock {\em J. Comput. Phys.}, 400(108877):108877, January 2020.

\bibitem{jing2025convergence}
Yang Jing and Lei Li.
\newblock {Convergence Analysis of OT-Flow for Sample Generation}.
\newblock {\em Numerical Mathematics: Theory, Methods and Applications}, 18(2):325--352, 2025.

\bibitem{JKO1998}
Richard Jordan, David Kinderlehrer, and Felix Otto.
\newblock {The Variational Formulation of the Fokker--Planck Equation}.
\newblock {\em SIAM Journal on Mathematical Analysis}, 29(1):1--17, 1998.

\bibitem{Kac1956-hi}
M~Kac.
\newblock Foundations of kinetic theory.
\newblock In {\em {Proceedings of the Third Berkeley Symposium on Mathematical Statistics and Probability}}, volume~3, pages 171--197. University of California Press Berkeley, Los Angeles, California, 1956.

\bibitem{keller1971model}
Evelyn~F Keller and Lee~A Segel.
\newblock Model for chemotaxis.
\newblock {\em Journal of theoretical biology}, 30(2):225--234, 1971.

\bibitem{Kolodziejczyk2025kc}
Martin Kolodziejczyk, Michela Ottobre, and Gideon Simpson.
\newblock {Counting the number of stationary solutions of partial differential equations via infinite dimensional sampling.}
\newblock {\em Philosophical transactions. Series A, Mathematical, physical, and engineering sciences}, 383(2298):20240239, June 2025.

\bibitem{kuramoto1981rhythms}
Yoshiki Kuramoto.
\newblock Rhythms and turbulence in populations of chemical oscillators.
\newblock {\em Physica A: Statistical Mechanics and its Applications}, 106(1-2):128--143, 1981.

\bibitem{li2025solving}
Lei Li, Yijia Tang, and Jingtong Zhang.
\newblock {Solving Stationary Nonlinear Fokker--Planck Equations via Sampling}.
\newblock {\em SIAM Journal on Applied Mathematics}, 85(1):249--277, 2025.

\bibitem{li2020random}
Lei Li, Zhenli Xu, and Yue Zhao.
\newblock {A random-batch Monte Carlo method for many-body systems with singular kernels}.
\newblock {\em SIAM Journal on Scientific Computing}, 42(3):A1486--A1509, 2020.

\bibitem{xiang2004}
Tiejun Li, Pingwen Zhang, and Xiang Zhou.
\newblock {Analysis of 1 + 1 Dimensional Stochastic Models of Liquid Crystal Polymer Flows}.
\newblock {\em Comm. Math. Sci.}, 2(2):295--316, 2004.

\bibitem{lin2022computing}
Bo~Lin, Qianxiao Li, and Weiqing Ren.
\newblock Computing the invariant distribution of randomly perturbed dynamical systems using deep learning.
\newblock {\em Journal of Scientific Computing}, 91(3):77, 2022.

\bibitem{lu2024score}
Jianfeng Lu, Yue Wu, and Yang Xiang.
\newblock {Score-based transport modeling for mean-field Fokker-Planck equations}.
\newblock {\em Journal of Computational Physics}, 503:112859, 2024.

\bibitem{McKean1966}
H.~P. {McKean}.
\newblock {A class of Markov processes associated with nonlinear parabolic equations}.
\newblock {\em Proceedings of the National Academy of Sciences}, 56(6):1907--1911, 1966.

\bibitem{McKean1967}
H.~P. {McKean}.
\newblock Propagation of chaos for a class of nonlinear parabolic equations.
\newblock {\em Lecture Series in Differential Equations}, 7:41--57, 1967.

\bibitem{monmarche2025long}
Pierre Monmarch{\'e}.
\newblock Long-time propagation of chaos and exit times for metastable mean-field particle systems.
\newblock {\em arXiv preprint arXiv:2503.00157}, 2025.

\bibitem{OttingerBook}
H.C. {\"O}ttinger.
\newblock {\em {Stochastic Processes in Polymeric Fluids: Tools and Examples for Developing Simulation Algorithms}}.
\newblock Springer-Verlag, Berlin Heidelberg, 1996.

\bibitem{papamakarios2021normalizing}
George Papamakarios, Eric Nalisnick, Danilo~Jimenez Rezende, Shakir Mohamed, and Balaji Lakshminarayanan.
\newblock Normalizing flows for probabilistic modeling and inference.
\newblock {\em Journal of Machine Learning Research}, 22(57):1--64, 2021.

\bibitem{PaulEmmanuel2022}
Thierry {Paul} and Emmanuel {Tr{\'e}lat}.
\newblock {From microscopic to macroscopic scale equations: mean field, hydrodynamic and graph limits}.
\newblock {\em arXiv e-prints}, page arXiv:2209.08832, September 2022.

\bibitem{rezende2015variational}
Danilo Rezende and Shakir Mohamed.
\newblock Variational inference with normalizing flows.
\newblock In {\em International conference on machine learning}, pages 1530--1538. PMLR, 2015.

\bibitem{shen2024entropy}
Zebang Shen and Zhenfu Wang.
\newblock {Entropy-dissipation Informed Neural Network for McKean-Vlasov Type PDEs}.
\newblock {\em Advances in Neural Information Processing Systems}, 36, 2024.

\bibitem{Sznitman1991}
Alain-Sol Sznitman.
\newblock Topics in propagation of chaos.
\newblock In Paul-Louis Hennequin, editor, {\em {Ecole d'Et{\'e} de Probabilit{\'e}s de Saint-Flour XIX --- 1989}}, volume 1464 of {\em Lecture Notes in Mathematics}, pages 165--251. Springer Berlin Heidelberg, 1991.

\bibitem{tang2022adaptive}
Kejun Tang, Xiaoliang Wan, and Qifeng Liao.
\newblock Adaptive deep density approximation for {Fokker-Planck} equations.
\newblock {\em Journal of Computational Physics}, 457:111080, 2022.

\bibitem{tugaut2014phase}
Julian Tugaut.
\newblock {Phase transitions of McKean--Vlasov processes in double-wells landscape}.
\newblock {\em Stochastics An International Journal of Probability and Stochastic Processes}, 86(2):257--284, 2014.

\bibitem{WANG2025107165}
Taorui Wang, Zheyuan Hu, Kenji Kawaguchi, Zhongqiang Zhang, and George~Em Karniadakis.
\newblock Tensor neural networks for high-dimensional {Fokker–Planck} equations.
\newblock {\em Neural Networks}, 185:107165, 2025.

\bibitem{zeng2023adaptive}
Li~Zeng, Xiaoliang Wan, and Tao Zhou.
\newblock Adaptive {Deep} {Density} {Approximation} for {Fractional} {Fokker--Planck} {Equations}.
\newblock {\em Journal of Scientific Computing}, 97(3):68, 2023.

\end{thebibliography}
\bibliographystyle{plain}

\end{document}